\documentclass[aps,prd,reprint,nofootinbib,groupedaddress]{revtex4-2}
\usepackage{graphicx}
\usepackage{dcolumn}
\usepackage{bm}
\usepackage{hyperref}
\usepackage{amssymb}
\renewcommand{\d}{\mbox{d}}

\newcommand{\intx}{\int d^4 x\,}

\newcommand{\al}{\alpha}
\newcommand{\bt}{\beta}
\newcommand{\gm}{\gamma}
\newcommand{\dl}{\delta}
\newcommand{\ep}{\epsilon}

\newcommand{\et}{\eta}
\newcommand{\kp}{\kappa}
\newcommand{\lm}{\lambda}
\newcommand{\rh}{\rho}

\newcommand{\ph}{\phi}
\newcommand{\ch}{\chi}
\newcommand{\ps}{\psi}

\newcommand{\Om}{\Omega}

\newcommand{\Lm}{\Lambda}

\newcommand{\Ph}{\Phi}

\newcommand{\half}{\frac{1}{2}}

\newcommand{\eela}[1]{\label{#1}\end{equation}}
\newcommand{\eeala}[1]{\label{#1}\end{eqnarray}}
\newcommand{\be}{\begin{equation}}
\newcommand{\ee}{\end{equation}}
\newcommand{\bea}{\begin{eqnarray}}
\newcommand{\eea}{\end{eqnarray}}

\newcommand{\ffit}{f_{a_G}}
\newcommand{\Ft}{\tilde F}
\newcommand{\Gtata}{G^t_{\;\,t}}
\newcommand{\Gtar}{G^t_{\;\,r}}
\newcommand{\Grr}{G^r_{\;\,r}}
\newcommand{\Gthth}{G^\theta_{\;\,\theta}}
\newcommand{\Gphph}{G^\ph_{\;\,\ph}}
\newcommand{\Gmumu}{G^\mu_{\;\,\mu}}
\newcommand{\Gmunu}{G^\mu_{\;\,\nu}}
\newcommand{\fgi}{j}
\newcommand{\fref}{\gf_{\rm p}}
\newcommand{\rref}{r_{\rm p}}
\newcommand{\etmax}{\et_{\rm max}}
\newcommand{\amax}{a_{\rm max}}
\newcommand{\Nav}{\langle N_4\rangle}
\newcommand{\st}{\tilde s}
\newcommand{\nt}{\tilde n}
\newcommand{\cGS}{c_{G{\rm S}}}
\newcommand{\cGR}{c_{G0}}
\newcommand{\GN}{G_{\rm N}}

\newcommand{\gf}{y}
\newcommand{\fg}{g}

\newcommand{\reff}{r_{0 m}}

\newcommand{\mO}{\mathcal{O}}

\newcommand{\rb}{r_{\rm b}}

\newcommand{\rmm}{r_{\rm m}}
\newcommand{\etm}{\et_{\rm m}}
\newcommand{\rms}{r_{\rm ms}}
\newcommand{\xms}{x_{\rm ms}}
\newcommand{\rmin}{r_{\rm min}}
\newcommand{\rmax}{r_{\rm max}}

\newcommand{\rnotcon}{r_{0{\rm c}}}
\newcommand{\rcon}{r_{\rm c}}
\newcommand{\acon}{a_{\rm c}}
\newcommand{\Vcon}{V_{\rm c}}

\newcommand{\ellt}{\tilde\ell}

\newcommand{\ds}{d_{\rm s}}

\begin{document}
\title{Using massless fields for observing black hole features in the collapsed phase of Euclidean dynamical triangulations}
\author{Jan Smit\\
Institute for Theoretical Physics, University of Amsterdam, \\
Science Park 904, P.O.Box 94485, 1090 GL Amsterdam, the Netherlands.
}

\begin{abstract}
We report on an old computation of propagators of massless scalar fields on an ensemble of configurations in 4D Euclidean dynamical triangulations in the collapsed (crumpled) phase. The resulting quantum average is used to construct the scale factor of a 4-D rotational invariant metric.
This new scale factor is non-zero at the origin, which we assume to be caused the presence of the well-known singular structure in the collapsed phase.  
The scale factor depends on an overall integration constant, which is determined by comparison with the implied volume at intermediate distances.
We construct a transformation to a 3-D rotational invariant metric, which reveals Euclidean black hole features at an instant in time, with a horizon separating interior and exterior parts. 
Effective Einstein equations in the presence of a `geometric condensate' are assumed, and computed with the software OGRe. Also explored briefly is a qualitatively different (not preferred) interpretation of the data which follows from comparison with an inversely transformed Euclidean regular black hole metric.
\end{abstract}

\maketitle

\section{Introduction}
\label{secintro}

In the dynamical triangulation approach to Euclidean quantum gravity (EDT) the configurations of the regulated path integral form a lattice consisting of equilateral four-simplexes `glued' together \cite{Agishtein:1991cv,Ambjorn:1991pq,Bilke:1998bn} (for reviewing texts see e.g.\ \cite{Thorleifsson:1998jr,Ambjorn:1999ix,Ambjorn:2013eha,Laiho:2016nlp}).
A scenario for a continuum limit has been proposed  \cite{Laiho:2016nlp} for a model with three parameters, two of which monitor the (bare) Newton and cosmological couplings, $G_{{\rm N}0}$ and $\Lm_0$, the third controls a `measure term'. This measure term is a local factor in the summation measure over configurations  \cite{Bilke:1997sc,Bilke:1998vj} which can be written as a term in the action that depends logarithmically on the Regge curvature \cite{Ambjorn:1999ix,Coumbe:2014nea} (initially a different one was introduced in \cite{Bruegmann:1992jk}, recently revisited in \cite{Asaduzzaman:2022kxz}).

A customary notation uses dimensionless parameters $\kp_2\propto 1/G_{{\rm N}0}$ and $\kp_4\leftrightarrow \Lm_0$, and $\bt$ for the measure term.
Numerical simulations have revealed two phases, a `collapsed phase' (a.k.a.\ `crumpled phase') at strong coupling and an `elongated phase' at weak coupling, separated by a first-order transition line in the $\kp_2$-$\bt$ phase diagram \cite{Bialas:1996wu,deBakker:1996zx,Ambjorn:2013eha,Rindlisbacher:2015ewa,Coumbe:2014nea}. In the scenario of  \cite{Laiho:2016nlp} the continuum limit is to be approached along the phase boundary on the collapsed side towards weaker couplings, i.e.\ larger $\kp_2$.

Similar phases appear in the formulation using well-defined time slices called `causal dynamical triangulation' (CDT) \cite{Ambjorn:2012jv,Loll:2019rdj,Ambjorn:2022naa}, and approaches to a possible UV fixed point along `lines of constant physics' are described in \cite{Ambjorn:2014gsa,Ambjorn:2016cpa,Gizbert-Studnicki:2023usv}.

A continuum limit towards a phase boundary from the strong coupling side is also proposed in a different formulation with variable lattice edge lengths \cite{Hamber:2015jja,Rocek:1981ama}.

The original EDT formulation had no measure term, i.e.\ $\bt=0$  \cite{Agishtein:1991cv,Ambjorn:1991pq}.
Numerical simulations at $\bt=0$ revealed that the two phases have very distinct properties. The elongated phase is dominated by baby universes with branched polymer characteristics \cite{Ambjorn:1995dj}. Deep in the collapsed phase a `singular structure' appears consisting of two adjacent `singular vertices' that belong to a macroscopic number of four-simplices that grows linearly with the lattice size \cite{Hotta:1995ud,Hotta:1995ca,Catterall:1995ig,Catterall:1997xj,Bialas:1996eh}. Going towards weaker coupling the structure breaks up leaving only one singular vertex which finally disappears at the phase boundary \cite{Catterall:1997xj}.

Visualizations of typical configurations at $\bt=0$ show  fractal tree-like `branched-polymer' structures in the elongated phase and a dense `blob' with trees sticking out in the collapsed phase \cite{Ambjorn:2013eha,Rindlisbacher:2015ewa,Laiho:2016nlp}.
Branched polymers have been linked to conformal fields \cite{Hikami:2017sbg,Poland:2018epd} and one can imagine the blob as representing a naked singularity or a black hole, and the trees representing a boundary field theory, similar to AdS/CFT \cite{Maldacena:1997re,Witten:1998qj,Witten:1998zw}.
At negative $\bt$, upon increasing $-\bt$ and $\kp_2$ the system enters a region called `crinkled' \cite{Bilke:1997sc,Bilke:1998vj} in which the blob stratifies and includes larger distances \cite{Ambjorn:2013eha,Coumbe:2014nea} with an accompanying smaller lattice spacing \cite{Laiho:2016nlp}.
These features suggest studying quantal properties of black-holes by simulations in EDT.

At $\bt=0$ there will be lattice artefacts in the results but they may still inform us about quantum-gravitational physics.
The great advantage of the numerical simulations is that they yield truly non-perturbative results which can qualitatively and even semi-quantitatively describe physics. We recall that lattice QCD at strong coupling (i.e.\ far from its continuum limit) gives a fair characterization of the low mass hadrons (as summarized for example in \cite{Smit:2002ug}, section 7.4).

In \cite{deBakker:1996qf,deBakker:1993dz} gravitational binding energy between two massive scalars was computed in EDT in the `quenched approximation'
(in perturbation theory this corresponds to leaving out closed particle loops).
A non-zero binding energy $E_{\rm b}$ came out, with a dependence on the particle masses $m$ that was not understood. A recent study \cite{Smit:2021lyq} suggested an explanation in terms of a blow-up effect in the bound state wave function causing finite-size squeezing of the system with a \emph{decreasing} ratio $E_{\rm b}/m$ with \emph{increasing} $m$, in a large mass region. The effect is proportional to $m\, G_{\rm N}$, with $G_{\rm N}$ a renormalized Newton coupling. In an elaborate analysis of the results in \cite{deBakker:1996qf} the renormalized Planck length $\sqrt{G_{\rm N}}$ was estimated to be fairly large ($\approx 8$) in lattice units, near the phase transition on the collapsed side. The relevance of the black hole scale $m\,G_{\rm N}$ came as a surprise. (This explanation needs confirmation on larger lattices.)

A controlled computation of the binding energy appeared in \cite{Dai:2021fqb}. This work used extrapolation to the continuum limit in the above mentioned scenario, scale-dependent dimensions allowing relatively small lattices and extrapolation to infinite volume, with the result $\sqrt{\GN}=3.9 \pm 0.7$ in lattice units.
A similar value was subsequently found using a completely different method \cite{Bassler:2021pzt}.
These computations employed a different ensemble of so-called degenerate triangulations \cite{Bilke:1998bn}.
(In \cite{Smit:2021lyq} the analysis uses dimension 4 which requires avoiding short distances.)

In \cite{deBakker:1994zf} effective curvature observables based on a `volume-distance correlator' were computed which showed at short distances negative curvature in the collapsed phase and positive curvature in the elongated phase.\footnote{Negative curvature was also found in the strong coupling phase of the formulation with variable lattice edge lengths \cite{Hamber:2015jja}.} 
This was refined further in \cite{Smit:2013wua} into effective \emph{metrics} describing an average geometry.

The present work focuses on the collapsed phase \emph{away} from the phase transition.
We present results for the propagators of massless scalar-fields
obtained from simulations in 1996 by B.V.\ de Bakker in collaboration with the present author.
The negative curvature in the collapsed phase found earlier raised the question whether this would be `confirmed' by an exponential decay of massless propagators at large distances, as in hyperbolic space in the continuum.
Evidence for such decay was indeed found and briefly mentioned in \cite{Smit:1996yq}, but for logistic reasons the write-up got stuck at the stage of `a draft of a draft' \cite{deBakker:1996unpub}.
Part of these results could be recovered recently. They are used here to derive, in a fundamentally different way, metrics for average geometries as observed through the propagators.

The propagator reacts to the quantum geometries in a way that fundamentally differs from the volume-distance observable.
A clue to an explanation may be found in the plot in figure 2.7 of \cite{deBakker:1995yb}, which shows that a massless scalar field propagator on a configuration deep in the {\em elongated phase} becomes constant on the thin branches of its tree-like structure. It is reminiscent of the constancy of the electric potential in a Faraday cage. 

Our knowledge of distances is based primarily on electrodynamics involving the massless photon, and on the Planck scale all fields in the Standard Model are practically massless. 

The above mentioned volume-distance correlator is the average three-volume at a fixed distance from an arbitrary origin. With lattices of $S^4$ topology it leads naturally to isotropic average metrics of the form
\be
ds^2 =d\et^2 + a(\et)^2\, d\Om_3^2\,,
\label{isotropic}
\ee
with $\d\Om_3$ the line-element on the unit three-sphere and $\et$ a radial coordinate in four dimensions.

In the continuum, the scale factor $a(\et)$ is closely related to the derivative of the four-volume $V(\et)$ within distance $\et$ from the origin:
\be
V(\et ) = \int d\Om_3 \int_0^{\et } d\bar \et \, a(\bar \et )^3\,,\quad \partial_\et  V(\et ) = 2\pi^2 a(\et )^3\,.
\label{contV}
\ee
On the lattice, a volume-distance correlator $V^\prime(\et)$ corresponding to $\partial_\et V(\et)$ has been used to construct a lattice version\footnote{Examples are given in appendices \ref{appintroDT} and \ref{appEdSmetric}. \emph{Note that in appendices \ref{appintroDT}, \ref{appEdSmetric}, \ref{appmasslessprop} and \ref{appcompcur}, $\et$ is denoted by $r$ --- not to be confused with $r$ in (\ref{spherical}).} }
of $a(\et)$.

In the present work the important observable is the propagator $G(\et)$ of a massless scalar field. In the continuum it has to satisfy the Laplace-Beltrami equation for the metric (\ref{isotropic}),
\be
\left[\frac{d}{d\et} + \frac{3}{a(\et)}\frac{d\,a(\et)}{d\et} \right]\frac{d}{d\et}\, G(\et)=0\,, \quad \et>0\,,
\label{Ga}
\ee
with appropriate boundary conditions.
Given $G(\et)$, we can use (\ref{Ga}) as a first-order differential equation for a scale factor $a(\et)$.
The solution is 
\be
a(\et) = c_G\,a_G(\et)\,,\quad a_G(\et)=\left[-\frac{1}{\partial_\et G(\et)}\right]^{1/3}\,,
\label{introaG}
\ee
with $c_G$ an integration constant. Replacing the continuum $\partial_\et G(\et)$ by EDT data $G^\prime(\et)$ gives a new scale factor for the average quantum geometry.

The newly obtained scale factor $a(\et)$ determines how the three-volume described by the metric grows with the distance from the origin, which depends on the integration constant $c_G$.  We seek to determine $c_G$ by comparison with the scale factor based on the volume-distance correlator $V(\et)$. Both $G(\et)$ and $V(\et)$ involve a volume-average over the origin $\et=0$ and there will be branched-polymer contributions that are treated differently by the two observables. It turns out that at \emph{medium distances}\footnote{The lattice scale factor $\propto V^\prime(\et)^{1/3}$ (interpolated) has a maximum at $\etm$ and inflexion points $\et^\pm$ in $\et ^>_< \etm $. For clarity we define short, medium, transition and long distances respectively as $0< \et <\et^-$, $\et^- <\et < \etm$, $\etm < \et <\et^+$ and $\et^+ < \et$. }  the ratio of the volumes described by the two different scale factors has a region of slow variation, a stationary point at which it may be set to one and thereby fix $c_G$.

A major characteristic of this new scale factor is that it suggests fit functions that are non-vanishing at the origin. 
Their extrapolation to the origin should not suffer greatly from lattice artefacts or running dimensions and we assume this extrapolation to have physical significance.

To investigate wether the collapsed state has properties showing the presence of a black hole
we make a transformation to spherical co\"{o}rdinates in which such objects have been traditionally introduced.
The transformed line element has the form
\be
ds^2 = g_{tt}(r,t)dt^2 + 2g_{rt}(r,t)dr\, dt + g_{rr}(r,t)dr^2 + r^2\, d\Om_2^2 .
\label{spherical}
\ee
Here $d\Om_2$ is the line element on $S^2$ with unit radius, $t$ is a Euclidean time variable and $r$ a radial co\"{o}rdinate in three dimensions.

When the derivative $a^\prime(0)=0$, it turns out that at a time-instant $t=0$ a black hole appears with a Euclidean de Sitter (EdS) interior. It is interesting to also consider possibilities in which $a^\prime(\et)$ near the origin is non-zero but relatively small. Thus we are led to properties similar to
`regular black holes' (for the latter see e.g.\ \cite{Bonanno:2022rvo} and references therein).

Allowing also for a quantum mechanical `vacuum condensate' \cite{Chapline:2000en,Mazur:2001fv,Mazur:2004fk,Cattoen:2005he,Mazur:2015kia,Simpson:2019mud}, it makes sense to 
assume that the Einstein equations hold in the effective sense. This gives the form of the energy-momentum tensor of the condensate, $T^\mu_{\;\,\nu}$,
\be
G^\mu_{\;\,\nu} = 8 \pi \GN\, T^\mu_{\;\,\nu}\,,\quad G^\mu_{\;\,\nu} = R^\mu_{\;\,\nu} - \half\, R\, \dl^\mu_{\;\,\nu}\,,
\label{EinsteinT}
\ee
where $\GN$ is a renormalized Newton constant. For the calculation the software OGRe \cite{Shoshany:2021iuc} is very helpful.

The collapsed phase has been judged as being un-physical, presumably because of its singular vertices and
its Hausdorff dimensions substantially larger than four.
The occurrence of singular vertices has been well explained in terms of `balls-in-boxes' models 
(cf.\ \cite{Bialas:1997qs,Bialas:1999ad} and references therein).
Here they are interpreted physically as suggesting the presences of black holes or similar objects. A precise comparison of locations cannot be made because the available numerical data has been volume-averaged, blurring locations.

Hausdorff dimensions (local and global) have been estimated \cite{Ambjorn:1995dj,Coumbe:2014nea,Laiho:2016nlp,Asaduzzaman:2022kxz}
from the behavior of volume-volume and volume-distance correlators.
The latter were used in \cite{Smit:2013wua} to construct effective \emph{metrics} at short distances in \emph{four} dimensions.
When they have hyperbolic features, sampling them at discrete distances will easily produce values of local Hausdorff dimension observables that are larger than four. We see no need of these for a characterization of the crumpled phase. 

Global Hausdorff dimensions involve scaling at increasing volume, which may be accompanied by `anomalous dimensions'. Large values $\gg 4$ are found rather deep in the collapsed phase \cite{Coumbe:2014nea}, but towards the phase boundary
they diminish and become compatible with or close to four \cite{Laiho:2016nlp,Asaduzzaman:2022kxz}. In \cite{Smit:2013wua} high-dimensional or even logarithmic scaling was seen in a volume-distance correlator expressed in the lattice geodesic distance. However, observables expressed in so-called `continuum' distances (introduced in the same reference) may behave differently, since they involve a volume-dependent re-scaling. 

The remainder of this article is organised as follows. 
Section \ref{secaG} introduces our numerical results for $a_G(\et)$ in (\ref{introaG}). Data for $\kp_2$ farthest away from the phase transition are chosen for further analysis in terms of a rational fit function. Section \ref{seccG} describes our choice of the integration constant $c_G$. For comparison  other possibilities are also mentioned. In section \ref{secinittrans} we study transformations to the 3+1 D spherical co\"{o}rdinate system (\ref{spherical}). A numerically obtained regular transformation is described in subsection \ref{secnumerical}. It appears to have a singular limit as $t\to 0$ which can be obtained analytically from a singular transformation described in subsection \ref{secttozero}. The limiting metric has black hole features shown in section \ref{secbhfeat}. The corresponding Einstein tensor is also shown, highlighting energy density and pressure of the condensate. 

Section \ref{secH} deals with the question: given a regular black hole in the 3+1 spherical co\"{o}rdinate system, what is the result of the inverse transformation to the 4 D spherical system (\ref{isotropic})? With the knowledge obtained in section \ref{secinittrans} this question is answered for a specific example,  the Hayward model. Its qualitative properties might be relevant for distinguishing  regular from genuine black holes in simulations with larger volumes.
Concluding remarks are in section \ref{secconc}. 

Appendix \ref{appintroDT} summarises details of the EDT formulation and the volume-distance correlator.
In appendix \ref{appEdSmetric} we show that it can be fitted well by an EdS \emph{metric} at \emph{medium distances}. When expressed in terms of `continuum' distances the fit closely resembles the one in \cite{Laiho:2016nlp}. 
It is found to have good  scaling behavior. 

Appendix \ref{appmasslessprop} is a literal rendition of section 3 in the draft \cite{deBakker:1996unpub}. It contains exact expressions for the massless propagator $G(\et)$ in continuous space with constant negative curvature $-12/r_0^2$ and its exponential fall-off $\propto \exp(-3 \et/r_0)+ \cdots$ as $\et\to\infty$. Numerical results  are presented for the lattice $G(\et)$ fitted by the exponential form $a\,\exp(-m\,\et)+ c$. However, further analysis in appendix \ref{appcompcur} leads to the conclusion that the interpretation $m=3/r_0$ with $r_0$ a hyperbolic curvature radius is in-appropriate and misleading. (This does not affect the fact that the exponential fit also considered in section \ref{secaG} describes the $a_G(\et)$ data well in a limited fit domain.)

Appendices \ref{appET} and \ref{appEH} contain technical details of the co\"{o}rdinate transformation.

\section{Metric scale factor from measured propagator}
\label{secaG}

With the propagators $G(\et)$ from the numerical simulation described in appendix  \ref{appmasslessprop} the relation (\ref{Ga}) can be used to obtain an average metric scale factor as `seen' by the propagator, by treating (\ref{Ga}) as a differential equation for $a(\et)$:
\be
3\, \partial_\et\ln[a(\et)] = -\partial_\et \ln\left[-\partial_\et G(\et)\right]\,.
\ee
The solutions for $a(\et)$,
\be
a(\et) \propto \left[-\frac{1}{\partial_\et G(\et)}\right]^{1/3}\,,
\ee
have a multiplicative integration constant.
On the lattice we implement the discrete derivative as
\be
G'(\et+1/2) = G(\et+1)-G(\et)\,,
\label{Gprime}
\ee
and define a lattice version of the scale factor by
\be
a_G(\et) = \left[-\frac{1}{G'(\et)}\right]^{1/3}\,.
\label{aG}
\ee
The scale factor to represent the average geometry in the continuum contains the multiplicative integration constant $c_G$,
\be
a(\et) = c_G\,a_G(\et)\,.
\label{aGcont}
\ee

\begin{figure}
\includegraphics[width=8cm]{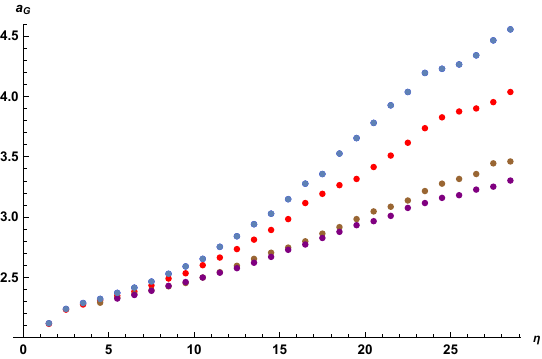} 
  \caption{
  Lattice $a_G(\et)$, top to bottom $\kp_2= 1.240$, 1.245, 1.250, 1.252; $N_4=32$k.
  }
\label{figaG}
\end{figure}

Results are available for $\kp_2=1.240$, 1.245, 1.250 and 1.252 at $N_4=32$ k. Figure \ref{figaG} shows $a_G(\et)$ for these $\kp_2$.
Unfortunately, we have no information about statistical errors of the retrieved data.

In this study of the collapsed phase it is important that $\kp_2$ is sufficiently far from the phase transition. Judging from the broad susceptibility peak in figure 14 of \cite{Smit:2013wua}, this holds for $\kp_2=1.240$, and to a lesser extent for $\kp_2=1.245$ which appears located at the edge of the peak (its maximum is at $\kp_2^{\rm c}=1.257$). But the two larger $\kp_2$ have penetrated the susceptibility peak and indeed, their points plotted in figure \ref{figaG} appear separated from the two lower $\kp_2$.
In the following we concentrate primarily on case $\kp_2=1.240$.

The first three data points at $\et=1.5$, 2.5 and 3.5, where all $\kp_2$ data merge visually, are presumably strongly subjected to lattice artefacts; discarding these, a mental extrapolation to $\et=0$ suggests a non-zero $a_G(0) \approx 2.2$.

The pennant-like feature of the last five data points ($\et\geq 24.5$) is present in both $\kp_2=1.240$ and 1.245 which suggests that it is not due to statistical fluctuations.  It may indicate a drop in the contributing volume near $\et=25$ and also a change in the character of contributing geometries. 
At large distances, $\et \gtrsim 25$,  we expect branched polymers to dominate the contributions to $a_G(\et)$
(cf.\ appendix \ref{appintroDT}).

\begin{figure}[t]
\includegraphics[width=8cm]{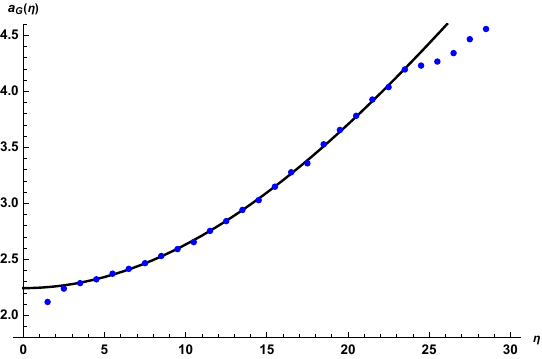} 
  \caption{
    Lattice $a_G(\et)$ data (blue dots) and fit function $f_{\rm rat}(\et)$ (black curve); 
$N_4=32$ k, $\kp_2=1.240$.
  }
\label{figFitsaG}
\end{figure}

A good choice of fit function and fit domain appears to be given by
\be
f_{\rm rat}(\et) =
\frac{p_0 + p_1\, \et^2}{1+q_1\, \et^2}\,,\quad 
\et=\{5.5, 6.5, \ldots, 23.5\}\,.
\label{frat}
\ee
It does significantly better than a mere quadratic fit $p_0 + p_1 \et^2$ in the region $\et\geq 19.5$, and even in $\et\leq 7.5$. 
The fitted parameters for the smaller two $\kp_2$ 
are given by
\be
\begin{array}{cccccccccc}
\kp_2&p_0&p_1&q_1&p_1/p_0-q_1&
\\
1.240&2.24&0.00447&0.000219&0.0018\\
1.245&2.23&0.00497&0.000616&0.0016
\end{array}
\label{fitrat}
\ee
Figure \ref{figFitsaG} shows the fit for the $\kp_2$ furthest from the phase transition.

For analytic treatment the simple form
\be
f_{\rm ch}(\et) = c_{\rm ch} \cosh(\et/r_0)
\label{fcosh}
\ee
is useful for suggesting the form of co\"{o}rdinate transformations in section \ref{secttozero}. 
Qualitatively $f_{\rm ch}(\et)$ is similar to $f_{\rm rat}(\et)$ in the fit domain and it approaches asymptotically a Euclidean Anti-de Sitter (EAdS) scale factor with length scale $r_0$. However, here $r_0$  monitors behavior of the data up to medium distances only.
For $\kp_2=1.240$ and the same fit domain as in (\ref{frat}) its two parameters come out as $c_{\rm ch} =2.32$, $r_0=22$.

\begin{figure}
\includegraphics[width=8cm]{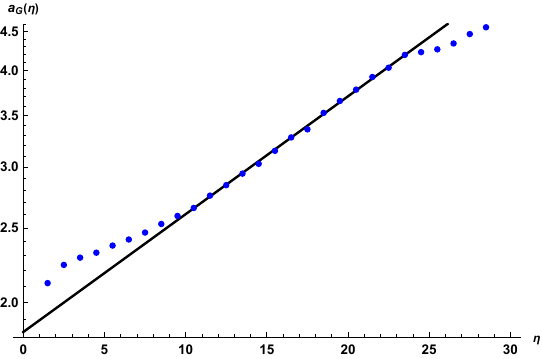} 
  \caption{
    Log-plot of $a_G(\et)$ data (dots) and fit function $f_{\rm e}(\et)=\exp(e_0+e_1 \et)$ (black line);  
    $\{e_0, e_1\} = \{0.605, 0.0353\}$; $N_4=32$k, $\kp_2=1.240$.
  }
\label{figFitaGl}
\end{figure}

For discussion we also show a log-plot in figure \ref{figFitaGl} of $a_G(\et)$ fitted by an exponential 
$f_{\rm e}(\et) = \exp(e_0 + e_1 \et)$ in the domain $\et \in \{9.5, \ldots, 24.5\} $. Because of the difference in fit domain the coefficients $e_{0,1}$ differ somewhat from the ones implied by the exponential fit made already for the $G(\et)$ data in \ref{parspropfit}. 

Of particular interest to our subsequent analysis turns out to be the region near the origin $\et=0$. To avoid strong lattice artefacts and variable dimension effects we should reach this region by extrapolation from $\et\gtrsim 6$ towards zero. In the following we \emph{define} the new metric through its fit function, including extrapolation to the origin, and choose the rational function:
\be
a_G(\et) = f_{\rm rat}(\et)\,
\label{defaG}
\ee
(keeping in mind that values $\et> 23.5$ are rather uncertain). 
It follows the trend of the data from $\et\simeq 10$ towards the origin better than $f_{\rm e}(\et)$, and should therefore also express better possible physics in the data. The fact that the derivative $f_{\rm rat}^\prime(\et)$ vanishes at the origin is relevant, we come back to this aspect in the Conclusions section \ref{secconc}.

\section{Determination of the integration constant}
\label{seccG}

\begin{figure}[t]
\includegraphics[width=8cm,angle=-0]{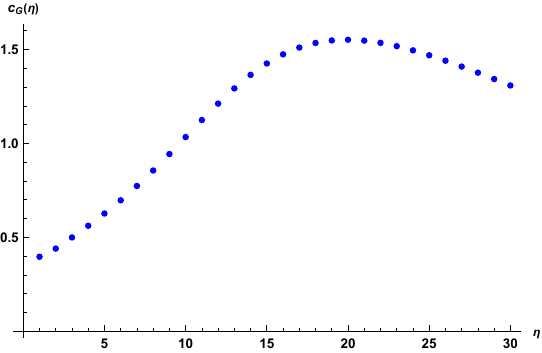} 
\caption{$c_G(\et)$ in (\ref{cGet});
$N_4=32$k, $\kp_2=1.240$.
}
\label{figcGet}
\end{figure}

The metric scale factor $a(\et)=c_G \, a_G(\et)$ is completed by specifying the integration constant $c_G$. 
According to $a(\et)$, the volume within distance $\et$ from the origin is
\be 
V_G(\et)=2\pi^2\, c_G^3  \int_0^\et d\bar \et\, a_G(\bar \et)^3\,.
\ee
This can be compared with the four-volume in a ball within distance $\et$ from the origin as given by the volume-distance correlator, $V(\et) = N(\et) \, v_4$. The question is, at which distance. Medium distances are favoured here since $V(r)$ shows EdS scaling at medium distances (cf.\ appendix \ref{appEdSmetric}) and is assumed to have physical significance in this region. This scaling, based on $V^\prime(\et)$, is subtle and involves a factor $\lm$ between lattice distances and distances dubbed `continuum'. However, integrated volumes are to be compared directly (cf.\ (\ref{Vac}), (\ref{VcV})).  An $\et$-dependent $c_G(\et)$,
\be
 c_G(\et)=\left[\frac{N(\et)\,v_4}{ 2\pi^2 \int_0^{\et} d\bar \et \, a_G(\bar \et)^3}\right]^{1/3}_{a_G(\bar \et)\to f_{\rm rat(\bar \et)}}\,,
  \label{cGet}
\ee
is shown figure \ref{figcGet}. 

\begin{figure}
\includegraphics[width=8cm]{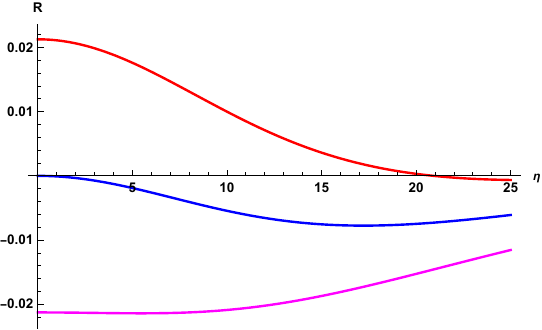} 
\caption{Curvatures $R(\et)$ for various $c_G$. Upper curve (red): $c_G=\cGS=5.3$; middle curve (blue): $c_G=\cGR=7.5$; lower curve (magenta): $c_G=\infty$.).
}
\label{figRS0inf2}
\end{figure}

It has a maximum the upper edge of the medium distance region, $\et\simeq 20$.
By a `principle of minimal sensitivity' we choose $c_G$ equal to the value at this point,
\be 
c_G= c_G(20)=1.55\,. 
\label{cG}
\ee
A similar value results from the following reasoning: We wish to avoid the region where branched-polymer `hair' contributions dominate over that of a `blob'. This is the long distance region which starts at $\et\simeq 25$. Choosing $\et=25$ gives $c_G(25)=1.46$. 

Other values are of interest to see the effect of $c_G$ on various observables, for example the curvature
\be
R(\et)=6\left[ -\frac{a^{\prime\prime}(\et)}{a(\et)} - \frac{a^{\prime}(\et)^2}{a(\et)^2} + \frac{1}{a(\et)^2}\right]\,.
\label{Ra12}
\ee
With $a(\et)=c_G\,f_{\rm rat}(\et)$ this becomes a rational function in which numerator and denominator are polynomials of the 4-th degree in $\et^2$. At the origin it simplifies to
\be
R(0) = \frac{6}{c_G^2\, p_0^2} - \frac{12\, p_1}{p_0} + 12\, q_1 \,.
\ee
Defining $c_G$ such, that $R$ vanishes at the origin leads to
\be
c_G=\frac{1}{\sqrt{2}\, p_0 \sqrt{p_1/p_0 - q_1}}\equiv\cGR\,,
\label{cGR0}
\ee
and $\cGR=7.5$.
A third value of $c_G$ will be considered later in section \ref{secbhfeat} when a comparing with the  Schwarzschild metric, in (\ref{cGS}): $\cGS=\sqrt{2}\,\cGR = 5.3$. 

The curvatures for $\cGR$ and $\cGS$ are shown in figure \ref{figRS0inf2}.
For our fiducial value (\ref{cG}) the curvature is positive and larger than for $\cGS$ by an order of magnitude (a factor $\gtrsim 25$ at $\et=0$). It is shown in figure \ref{figRvoldistEdS} together with curvature based on the volume-distance correlator.

\section{Transformation to spherical co\"{or}dinates}
\label{secinittrans}

We seek to transform the metric (\ref{isotropic}), more explicity given by
\bea
ds^2 &=&
d\et^2 + a(\et)^2\, d\ps^2 + a(\et)^2 \, \sin(\ps)^2 \, d\Om_2^2 \,,
\label{4D2}
\\
d\Om_2^2 &=& d\theta^2 + \sin(\theta)^2\,d\ph^2\,,
\eea
into the form (\ref{spherical}), shown again for convenience,
\be
ds^2 
 = g_{tt}(r,t)\,dt^2 + 2g_{rt}(r,t)\,dr\, dt + g_{rr}(r,t)\,dr^2
+ r^2\, d\Om_2^2 \,.
\label{spherical2}
\ee
The transformation is specified by a function $\gf(r,t)$ in
\be
r=a(\et)\,\sin(\ps)\,,\quad \gf(r,t)=a(\et)\,\cos(\ps)\,,
\ee
to be determined. In principle
\bea
0  &\leq&\et< \infty\,,\quad 0\leq\ps\leq\pi\,,
\\
 -\infty &<&t <\infty \,,\quad 0\leq r < \infty\,,
\eea
in practice $0\leq\et\leq\etmax$ with corresponding limits on $r$ and $t$ depending on $\gf(r,t)$.
Since $a>0$,
\bea
y &>& 0\,, \quad 0<\ps<\pi/2\,,
\\
y &<& 0\,, \quad \pi/2<\ps<\pi\,.
\eea
Note
\be
r^2 + \gf^2 = a^2\geq h^2\,,\quad h\equiv a(0)\,,
\label{rsqplusfsq}
\ee
which will be used in the following. We have 
\be 
h \leq a \leq \amax\,,\quad 0\leq r\leq \rb
\ee
where $\rb=\amax$, the maximal $r$ for $y=0$. With $\amax=a(\etmax)=c_G\, a_G(\etmax)$, $\etmax=23.5$ (the upper limit of the fit domain), and $c_G \simeq 1.5$ this gives 
\be 
\rb\approx 2\, h \,.
\label{rb}
\ee
The case with $a=h$ will be called `the boundary equation'.

In the region\footnote{This region will include $a\in(h,\amax)$.}  where $a(\et)$ is a monotonously increasing function of $\et$  it has a unique inverse $\et(a)$, and a positive function $F(a)$ can be defined by
\be
(d a/d \et)^2=F(a) = F\left(\sqrt{r^2 + \gf^2}\right).
\label{introF}
\ee
The new metric turns out as
\bea
g_{tt} &=& \frac{\dot \gf^2}{r^2 + \gf^2}\left(\frac{\gf^2}{F} + r^2\right)\,,
\label{gtt}
\\
g_{rt} &=& \frac{\dot \gf}{r^2 + \gf^2}\left(\frac{\gf(\gf \gf^\prime+r)}{F}-r(\gf-r \gf')\right)\,,
\label{grt}\\
g_{rr} &=&\frac{1}{r^2 + \gf^2}\left(\frac{(\gf \gf'+r)^2}{F}+(\gf-r \gf')^2\right)\,
\label{grr}
\eea
($\dot \gf = \partial_t \gf$, $\gf'=\partial_r \gf$).
We would like $\gf$ to be such, that the off-diagonal component of the metric vanishes,
\be
g_{rt}= 0\,.
\label{eqncross}
\ee
Equation (\ref{eqncross}) can be solved for $F$ to eliminate it from expression (\ref{grr}) for $g_{rr}$ and obtain
\be
g_{rr} = 1-r\,\gf^\prime/\gf\,,
\label{grrffp}
\ee
which is useful once $\gf$ that satisfies (\ref{eqncross}) is known.
Assuming that generically $\dot\gf \neq 0$, solving (\ref{eqncross}) for $\gf^\prime$ gives
\be
\gf^\prime= r \gf \frac{F-1}{\gf^2 + r^2 F}\equiv P\,,
\quad P=P(r,\gf)\,.
\label{fprime}
\ee
We also would like $\gf$ to satisfy the property
\be
g_{tt} = 1/g_{rr} \,,
\label{eqnggt}
\ee
as for the Schwarzschild Euclidean (Anti) de Sitter
(SE(A)dS) metrics \cite{Hawking:1982dh}. Requiring (\ref{eqnggt}) in addition to 
(\ref{eqncross}) leads to an equation for
$\dot\gf^2$, or
\be
\dot\gf = \pm\sqrt{F}\,.
\label{fdot}
\ee
The case with a minus-sign is the time-reversed version of the case with a plus-sign.
When using this equation in the following we choose the plus-sign to avoid double covering.

The differential $d\gf = P\, dr + \sqrt{F}\, dt$ is in general {\em imperfect},
i.e.\ loosely $\partial_t P \neq \partial_r \sqrt{F}$, meaning 
\bea
\partial_\gf P(r,\gf)\,\sqrt{F(\sqrt{r^2 + \gf^2})}
&\neq& \partial_r \sqrt{F(\sqrt{r^2 + \gf^2})}
\\
&& \left. + \partial_\gf \sqrt{F(\sqrt{r^2 + \gf^2})}\, P(r,\gf) \right. .
\nonumber
\eea
Requiring the differential to be perfect leads to a differential equation for $F$ which is easy to solve,
\be
dF/da=2(F-1)/a\; \Rightarrow\; F=1+ c_F\, a^2 = 1 + c_F (r^2 + \gf^2)\,,
\ee
where $c_F$ is an integration constant. Writing $c_F=\pm 1/r_0^2$ with $r_0>0$,
the solution of (\ref{fdot}$^+$) with initial condition $\gf(r,0)=0$ is,
\bea
\gf(r,t) &=&\sqrt{r_0^2 + r^2}\, \sinh\frac{t}{r_0}\,, \qquad\mbox{(EAdS)}
\label{fEADS}
\\
\gf(r,t) &=&\sqrt{r_0^2-r^2}\, \sin\frac{t}{r_0}\,, \quad\qquad\mbox{(EdS)}
\label{fEDS}
\eea
respectively for $c_F>0$ and $c_F< 0$. In the latter case $r$ is to be limited to $0<r<r_0$.
Upon substitution in (\ref{gtt}) --
(\ref{grr}) the time-dependence drops out and the diagonal EAdS and EdS metrics emerge,
\be
g_{tt}(r) = 1 \pm r^2/r_0^2\,.
\label{E(A)dS}
\ee
Furthermore, treating (\ref{introF}) as a differential equation for the scale factor $a(\et)$ leads for the $+$ case to the solution $a(\et)=r_0 \sinh[(\et-s_0)/r_0]$ with integration constant $s_0$ (or its $\et$-reversed version), corresponding to hyperbolic space.
In the $-$ case it leads to the spherical scale factor  $a(\et)=r_0 \sin[(\et-s_0)/r_0]$.

Integrating the imperfect differential $d\gf$ along a path in the $(r,\,t)$ plane gives a path-dependent result. To avoid this imperfection we shall in section \ref{secnumerical} release the condition $g_{tt}=1/g_{rr}$.

\subsection{Regular transformation to diagonal metric}
\label{secnumerical}

In this section the condition of diagonality, $g_{rt}=0$ is kept and the differential equation (\ref{fprime}) will be integrated at fixed times. 
Equation (\ref{fdot}) will be used only at one reference $r$ to obtain boundary conditions depending on time for the integration of (\ref{fprime}) along $r$. Then there is no imperfectness of the  co{\"o}rdinate transformation.

We concentrate on the fit function (\ref{frat}). It is convenient to rewrite its $a(\et)$ in the form
\be
a(\et) = h\,\frac{1+p\,\et^2}{1+q\,\et^2}\,,\quad
h=c_G\, p_0\,,\; p=\frac{p_1}{p_0}\,,\; q=q_1\,.
\label{apq}
\ee
The function $F(a)$ introduced in (\ref{introF}) can be determined via the inverse function $\et(a)$ of $a(\et)$; one finds
\bea
\et^2&=& \frac{a-h}{p h - q a},\,\quad
a^\prime(\et)^2 = \frac{4 h^2 (p-q)^2 \et^2}{(1+q\,\et^2)^4}\,
\label{eta}
\eea
and
\be
F(a)= \frac{4 q^3(a-h)(h p/q - a)^3}{h^2 (p-q)^2}\,.
\label{Frat}
\ee
The function $F(a)$ is positive between its zeros at $a=h$ and $a=h\,p/q$ with a maximum at $\amax$ somewhere in between, and only its monotonous branch in $h<a<\amax$ is to be used. 
In this section results will be shown with $c_G=1.55$ ($h\simeq 3.3$) (cf.\ (\ref{fitrat}) and (\ref{cG})), for which
$\amax$ is only slightly larger than $2h$.

Since an analytical treatment is already awkward in this case and might be prohibitive with more general fit functions, the following is a numerical exploration.

We choose a reference distance $\rref$ and erect a `time pole' $\fref(t;\rref)$ by solving $\dot \gf=\sqrt{F}$ numerically at $r=\rref$:
\be
\dot \fref(t;\rref)=\sqrt{F\left(\sqrt{\rref^2 + \fref(t;\rref)^2}\right)}\,,
\label{eqnpole}
\ee
with initial conditions $\gf=\gf_0$ at $t=t_0$ chosen as follows.
Guided by (\ref{fEADS}) we assume that $\gf>0$ ($\gf<0$) when $t>0$ ($t<0$). Compatible with this, the initial condition can be taken a minimal $|\gf|$ at $t=0$: for $\rref >h$, $\gf_0=0$, for $\rref<h$  the minimal $|\gf|$ has to comply with the boundary equation in (\ref{rsqplusfsq}):
\bea
\gf_0 &=& 0\,,\qquad\quad\quad  h \leq \rref \leq\rmax\,, \quad t_0 =0\,,
\nonumber\\
&=& \pm \sqrt{h^2-\rref^2}\,,\quad 0\leq \rref \leq h\,,\quad t_0 \to 0^\pm\,.
\label{initialcond}
\eea
The resulting $\fref$ is a monotonically rising function of $t$.
Next, equation  (\ref{fprime}) implementing $g_{rt}=0$ is solved (numerically) with a boundary condition attaching $\gf$ to the pole:
\be
\gf^\prime(r,t;\rref)=P(r,\gf(r,t;\rref))\,,\quad \gf(\rref,t;\rref) = \fref(t;\rref)\,.
\label{attpole}
\ee
The solutions $\gf(r,t;\rref)$ map the $(r,t)$ plane to the $(r,\gf)$ plane and the give a foliation of the latter. The foliation lines do not cross.

Consider erecting a second time pole
$\bar\fref(\bar t;\bar\rref)$
at a different position $\bar\rref\neq \rref$. This pole will be crossed by a foliation line $\gf(r,t;\rref)$ of the first pole,
$\gf(\bar\rref,t;\rref)=\bar\fref(\bar t;\bar\rref)$, which determines $\bar t$ in terms of $t$. This can be interpreted as a co{\"o}rdinate transformation of the time variable only:
$\bar t=\bar t(t)$.
The second pole leads to a second foliation
$\gf(r,\bar t;\bar\rref)$. However, substituting $\bar t=\bar t(t)$ will give just the original foliation.

In other words: results of different choices of the pole position are related by  co{\"o}rdinate time transformations that depend only on time. We shall see evidence of this in tensor components that transform as a scalar field under such transformations, in particular
the spatial components of the metric (which is diagonal in this section)
and the diagonal up-down components of the Einstein tensor, $G^\mu_{\;\,\mu}(r,t)$ 
(no summation).

Results will be shown for two choices of $\rref$ \& initial conditions for $\fref(t;\rref)$:
\bea
\rref &=& 2 h \; \&\; t_0=0\,,\;\fref(0;2 h)= 0\,,
\label{condpole}
\\
\rref &=&0\; \&  \;\fref(t_0;0)= \pm h\,(1 + c_t t_0^2)\;,\;
c_t=p-q\,,
\label{condpolein}
\eea
with very small but nonzero $c_t t_0^2>0$ to get the numerical integration started. (With the indicated choice of $c_t$, the solution of (\ref{eqnpole}) at small times is $\pm h(1+ c_t\, t^2 + \mO(t^4))$, cf.\ section \ref{secttozero}.)
The case of $-h$ in (\ref{condpolein}) corresponds to negative $t$, $\fref(t;\rref)$ and $\gf(r,t;\rref)$.

\begin{figure}
\includegraphics[width=8cm]{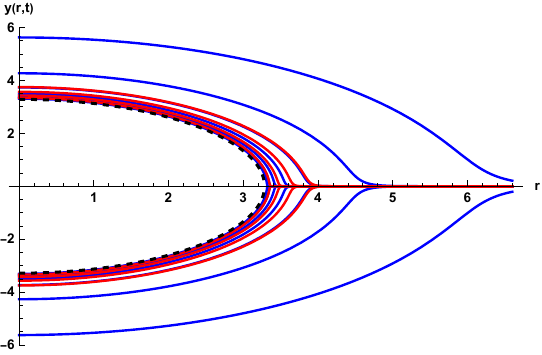} 
\caption{
Foliation curves $\gf(r,t)$ for $\rref= 2h$ and $t= \pm 10^{k-24}$, $k=0,4,8,\cdots,24$ (blue), and for $\rref=0$ at times $t= \pm \{3,5,7,9\}$ (red, partially overlaying the blues curves). The inner curve (dashed, black) represents $\pm\sqrt{h^2 - r^2}$; $h=3.3$ .
}
\label{figfol}
\end{figure}

\begin{figure}
\includegraphics[width=8cm]{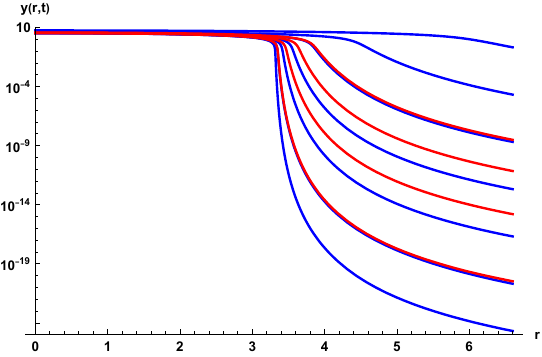} 
\caption{
Logarithmic plot corresponding to the positive $\gf$ part of figure \ref{figfol}.
}
\label{figfollog}
\end{figure}

Figure \ref{figfol} shows foliation curves obtained with the two choices of $\rref$ in (\ref{condpole}), (\ref{condpolein}). The curves run over the whole domain $0<r<\rmax$ and approach the same limit (dashed, black) when $t\to 0$.
Figure \ref{figfollog} shows a corresponding logarithmic plot.

The calculation of the metric needs derivatives of $\gf(r,t)$. Spatial derivatives can be expressed as functions of $\gf$ without derivatives using the first equation in (\ref{attpole}), repeatedly as needed, for example
\be
\gf^{\prime\prime} = \partial_r P(r,\gf) + \partial_y P(r,\gf)\, P(r,\gf)\,.
\ee
Time derivatives of $\gf$ were calculated using nearby foliations at $t\pm \ep$ and
\bea
\dot\gf(r,t)&\simeq& [\gf(r,t+\ep)-\gf(r,t-\ep)]/(2\ep)\,,
\label{dotsy}
\\
\ddot \gf(r,t)&\simeq& [\gf(r,t+\ep)+\gf(r,t-\ep)-2\,\gf(r,t)]/\ep^2\,,
\nonumber
\eea
with small $\ep$ typically of order $t/10$.
(Evaluating $\gf$ also at $t\pm 2 \ep$ one can approximate $\dddot\gf$ and also improve the above approximations. The first derivative $\dot\gf$ is needed for $g_{tt}$, but it turns out that the time derivatives cancel out of components of the Einstein tensor.)

\begin{figure}
\includegraphics[width=8cm]{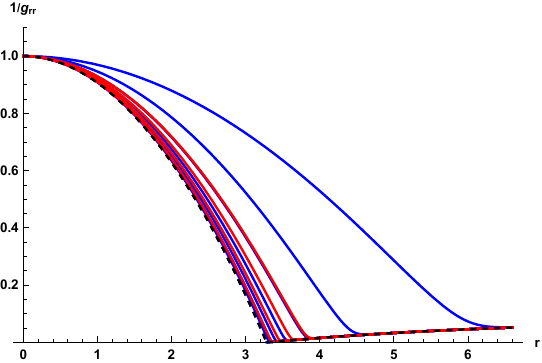} 
\caption{
Plot of $1/g_{rr}$. The dashed (black) curves represent $F(r)$ in the exterior and $1-r^2/h^2$ in the interior.
}
\label{figginvall}
\end{figure}

\begin{figure}
\includegraphics[width=8cm]{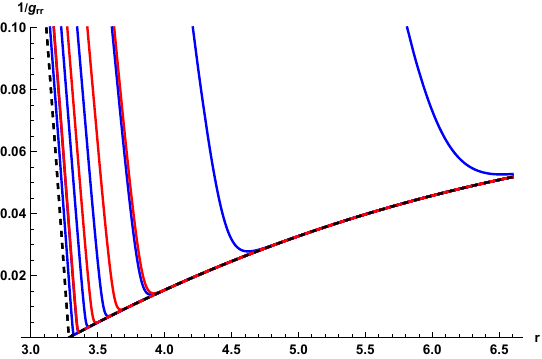} 
\caption{
Closeup of figure \ref{figginvall}. The envelope covers visually the $1/g_{rr}$ curves in the exterior region; it is zero at $r=h$.
}
\label{figginv}
\end{figure}

Figure \ref{figginvall} shows $1/g_{rr}(r,t)$ obtained with the help of (\ref{grrffp}). As $t$ approaches zero an envelope develops, which is represented by the dashed black curves. Figure \ref{figginv} shows a closeup.
The envelope is $F(r)$ in the exterior region $r>h$ and switches to $1-r^2/h^2$ in the interior. In the exterior this can be understood from taking the limit $\gf\to 0 $ and $\gf' \to 0$ in the basic form (\ref{grr}) for $g_{rr}$. In the interior the limit $\gf\to \pm \sqrt{h^2-r^2}$ leads to $F(h)$ in the denominator in (\ref{grr}) and one has to heed the fact that $F(h)=0$ (cf.\ section \ref{secttozero}).

\begin{figure}
\includegraphics[width=8cm]{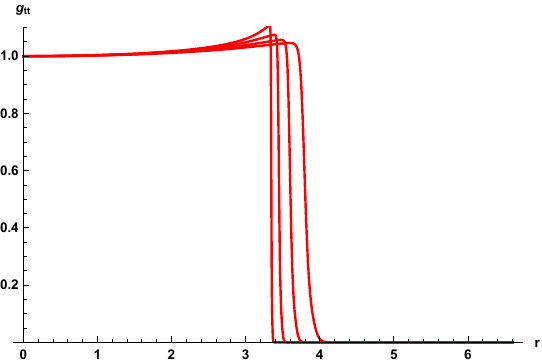} 
\caption{
Time component $g_{tt}(r,t)$ for $\rref=0$.
Left to right in $3.3<r<4$: $t=3$, 5, 7, 9.
}
\label{figgtt}
\end{figure}

\begin{figure}
\includegraphics[width=8cm]{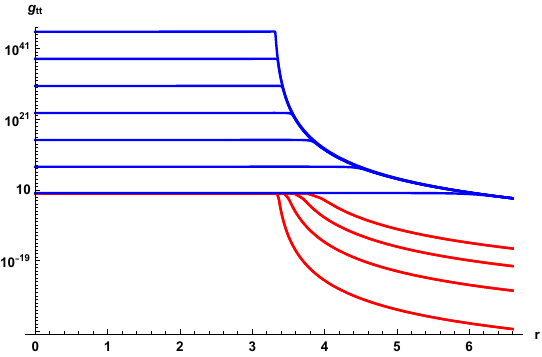} 
\caption{
Log-plot of $g_{tt}(r,t)$. The lower (red) curves correspond again to $\rref=0$, the upper (blue) curves to $\rref=2h$.
}
\label{figgttlog}
\end{figure}

The time component of the metric, $g_{tt}$ calculated from (\ref{gtt}) differs very much from $1/g_{rr}$ as shown in figures \ref{figgtt} and \ref{figgttlog}, note the vertical scale in the latter. The non-invariance of $g_{tt}$ under transformations of the time variable,  $t\to \bar t(t)$, has drastic effects here.

For the calculation of the components of the Einstein tensor the expressions given in Exercise 14.16 of \cite{Misner:1973prb} can be used; transformed to the Euclidean case they are recorded in appendix \ref{appET}.
Figure \ref{figpGtt} shows $\Gtata$ with the same conventions as in figures \ref{figfol} and following. It illustrates that $\rref=0$ and $\rref=2h$ give indeed the same $\Gtata$ curves as expected from the scalar nature of $\Gtata(r,t)$ under transformations $t\to \bar t(t)$. An envelope develops as $t\to 0$ which has a discontinuous jump at $r=h$. Figure \ref{figpGrr} shows similar phenomena in $\Grr$. For $\Gthth$ in figure \ref{figpGthth} the envelope is continuous ($\Gphph=\Gthth$). 

\begin{figure}
\includegraphics[width=8cm]{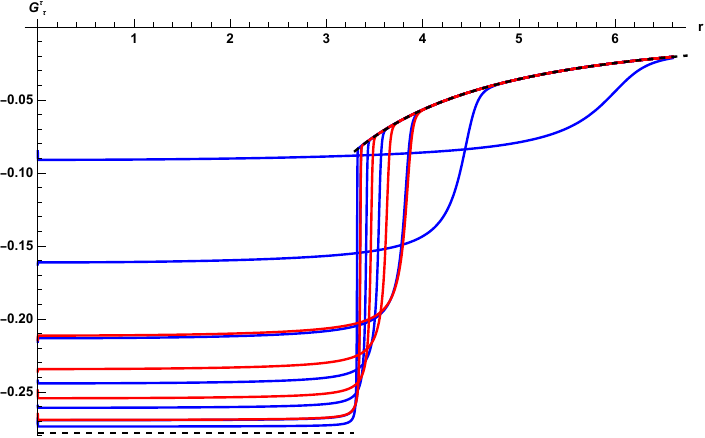} 
\caption{
$\Gtata$. Upper curves (blue) in $r<h$ correspond to $\rref=2h$, the curves occasionally joining (red) correspond to $
\rref=0$. The black-dashed lines represent (\ref{Gtt}), (\ref{Gttin}).
}
\label{figpGtt}
\end{figure}

 \begin{figure}
\includegraphics[width=8cm]{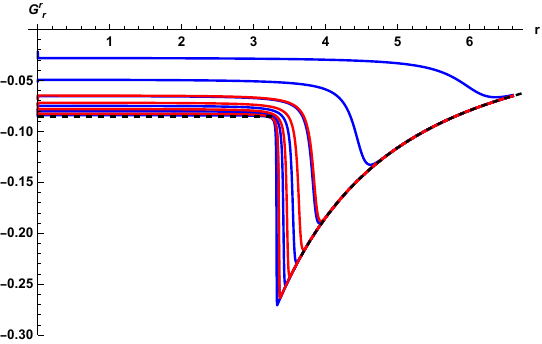} 
\caption{
$\Grr$. The black-dashed lines represent (\ref{Grr}), (\ref{Grrin}).
}
\label{figpGrr}
\end{figure}

\begin{figure}
\includegraphics[width=8cm]{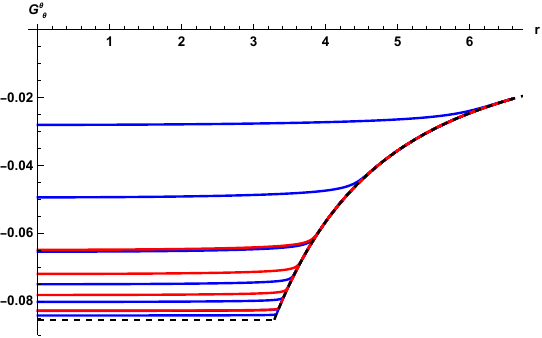} 
\caption{
$\Gthth$.  The black-dashed lines represent (\ref{Gtt}), (\ref{Grrin}).
}
\label{figpGthth}
\end{figure}

\begin{figure}
\includegraphics[width=8cm]{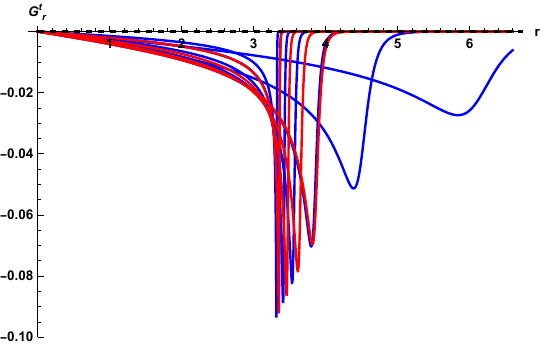} 
\caption{
$\Gtar$. 
}
\label{figpGtr}
\end{figure}

The Einstein tensor has an off-diagonal component $\Gtar$ shown in figure \ref{figpGtr}, which vanishes in the limit $t\to 0$ for $r\ne h$ and, as argued in appendix \ref{appEH}, also at $r=h$ when interpreted in the distributional sense.

The curvature $R$ provides a check on the calculations. The trace of the Einstein tensor is $-R$. Calculating $R=6(-a^{\prime\prime}/a - a^{\prime 2}/a^2 + 1/a^2)$ from the metric (\ref{isotropic}), which is a function of $\et^2$,
using $\et^2(a)$ in (\ref{eta}) with $a=\sqrt{r^2+ \gf^2}$ and the numerically calculated $\gf(r,t)$ to transform it to $(r,t)$ variables, we can compare it with $-\Gmumu(r,t)$. The two $R$ match accurately with a point-wise precision of order $10^{-14}$ \% (using Mathematica with default working conditions). This is surprising, since one expects from (\ref{dotsy}) that the time derivatives of $\gf$ will have only order percent accuracy. The reason is that $\dot\gf$ and $\ddot\gf$ cancel out of the Einstein tensor, cf.\ appendix \ref{appET}. The shape of $\Gmumu$ is similar to that of $\Gthth$.

\subsection{Singular transformation implementing $g_{tt}=g^{rr}$ at $t=0$}
\label{secttozero}

The numerical integration of $\gf'= P$ in section \ref{secnumerical} yielded diagonal metrics with spatial inverse components $1/g_{rr}=g^{rr}$ that approached a robust envelope when $t\to 0$.
On the contrary, the magnitude of the component $g_{tt}$ was highly sensitive to differing choices of boundary conditions (the two pole choices)
and the shape of $g_{tt}(r,t)$ differed from $g^{rr}(r,t)$.

In this section we seek to obtain $g_{tt}=g^{rr}$ in the limit $t \to 0$. We give up diagonality of the metric for general times but seek to recover it at zero time..
Consider erecting a pole as in the previous section but here at every $r\in(0,\rmax)$.  We wish to solve $\dot\gf = \sqrt{F}$ analytically at small times and try $\gf(r,t) = \fref(t;r)$.
It is helpful to take a brief look at the solution of this equation in case of the cosh-model,
\be
a(\et) = h \cosh(\et/r_0)\, ,
\label{coshmodel2}
\ee
which is simpler to solve than the rat-model.
(We recall that here $r_0$ is meant to parametrize primarily the small to intermediate distance region, $1/r_0^2$ may be thought to represent $2(p-q)$ of the rat-model to $\mO(\et^2)$.)
With $a^\prime(\et)^2= (h^2/r_0^2) (\cosh^2(\et/r_0^2) -1)$ the function $F$ in (\ref{introF}) turns out as
\be
F(\sqrt{r^2 + \gf^2})=( r^2 + \gf^2 -h^2)/r_0^2 \,.
\label{Fcosh}
\ee
The solution of (\ref{fdot}$^+$) with initial condition (\ref{initialcond})$|_{\rref\to r}$ is
\bea
\gf(r,t) &=& \pm \sqrt{h^2-r^2}\,\cosh(t/r_0)\,,\; 0< r < h\,,\;
t \stackrel{\textstyle >}{<} 0\,,
\nonumber 
\\&=& \sqrt{r^2-h^2}\,\sinh(t/r_0)\,,\quad\;\;\; r> h\,.
\label{fsln}
\eea
At $t=0$ this solution satisfies $\gf'=P$ (trivially in the exterior as $0=0$), and the resulting metric becomes indeed diagonal, $g_{rt}=0$, with
\bea
g_{tt}(r,0) = 1/g_{rr}(r,0) &=& 1-r^2/h^2\,,\quad r<h\,,
\label{gt0}
\\
&=& (r^2 - h^2)/r_0^2= F(r)\,,\quad r>h\,.
\nonumber
\eea

Generalizing to other models, e.g.\ the rat-model,  consider a factorized form as suggested by  (\ref{fsln}) in the exterior,
\be
\gf(r,t) = u(r)\, t\,.
\label{yut}
\ee
Equation
(\ref{grt}) shows that it leads to a vanishing off-diagonal component at $t=0$, $g_{rt}(r,0)=0$, independent of $u(r)$.
(The equation $\gf'= P$ is again solved trivially as $0=0$.)
Furthermore (\ref{grr}) gives
\be
g_{rr}(r,t)= \frac{1}{F(r)} + \mO(t^2)\,,
\label{grrF}
\ee
also independent of $u(r)$, whereas (\ref{gtt}) gives
\be
g_{tt}(r,t) = u(r)^2 + \mO(t^2)\,.
\label{gtatauout}
\ee
Hence, requiring $g_{tt}(r,0)\, g_{rr}(r,0)=1$ we get
\be
u(r)=\sqrt{F(r)}\,.
\label{uout}
\ee
and the equation $\dot \gf = \sqrt{F}$ is indeed satisfied at $t=0$.

In the interior the Ansatz is,
\be
\gf(r,t) =\pm \sqrt{h^2 - r^2}\,(1+c_t\, t^2)\,,\quad
t \stackrel{\textstyle >}{<} 0\,,
\label{yutin}
\ee
which has a time-dependence similar to that of the cosh-model for small $t$
(cf.\ first line in (\ref{fsln})). Since $F[h]=a'(0)^2=0$ (cf.\ (\ref{introF})), $F$ vanishes in the interior at $t=0$:
\be
F(\sqrt{r^2 +\gf(r,0)^2})=F[h]=0\,.
\ee
Its expansion in $t$ starts out as
\be
F(\sqrt{r^2 +\gf(r,t)^2})=(h-r^2/h)\,F'(h)\, c_t\, t^2 + \mO(t^4)\,.
\ee
The equation $\gf'= P$ is satisfied at $t=0$. When $t\to 0$, the off-diagonal part of the metric $g_{rt}$ in (\ref{grt}) contains $1/F$ which blows up, and it contains $\dot\gf$ which vanishes. Working out the details we find
\be
g_{rt}= - 2 r \left(1+\frac{2}{h F'(h)}\right)\,c_t\, t + \mO(t^3)\,.
\ee
Hence, the interior metric becomes also diagonal at time zero. Furthermore, after some algebra:
\bea
g_{rr}(r,t) &=& \frac{h^2}{h^2-r^2} +\mO(t^2)\,,
\label{grrin}
\\
g_{tt}(r,t) &=& \frac{h^2-r^2}{h^2}\, \frac{2 h\, c_t}{F'(h)} + \mO(t^2)\,.
\label{gtatactin}
\eea
Requiring $g_{tt}(r,0)\,g_{rr}(r,0) =1$ determines $c_t$,
\be
c_t=\frac{F'(h)}{2 h}\,,
\label{ctF}
\ee
and with this choice the equation $\dot\gf=\sqrt{F}$ is indeed solved to leading order in $t$.
For the rat-model this becomes
\be
c_t = p-q\,,
\label{ctpq}
\ee
which was used in section \ref{secnumerical} for the case $\rref=0$.

The metric component $g_{tt}(r,0)$ following from (\ref{Frat}), (\ref{gtatauout}), (\ref{uout}), (\ref{grrin}) -- (\ref{ctF}), is shown in figures \ref{figginvall} and \ref{figginv} by the dashed black curves.

The Einstein tensor $\Gmunu$ contains time derivatives of the metric which is non-diagonal at non-zero $t$ and a little complicated. For the daunting task of its calculation the software OGRe \cite{Shoshany:2021iuc} is very helpful. The results in the limit $t\to 0$ are fairly simple,
$\Gmunu(r,0)$ is diagonal and in the exterior region, $r>h$:
\begin{widetext}
\bea
\Gtata = G^\theta_{\;\,\theta}=G^\ph_{\;\,\ph} &=&
-\frac{1}{r^2}\,\left[1 +
\frac{4 (h p - q r)^2 (h^2 p - 2 h(p + 2 q )\,r + 5 q\, r^2)}{h^2 (p - q )^2}\right]\,,
\label{Gtt}\\
G^r_{\;\, r} &=& -\frac{3}{r^2} \left[1 +
\frac{4 (h - r)(h p - q r)^3}{h^2 (p - q )^2}\right]\,.
\label{Grr}
\eea
\end{widetext}
In the interior region $0<r<h$:
\bea
\Gtata&=& -\frac{3}{h^2}\,,
\label{Gttin}
\\
G^r_{\;\, r} = G^{\ph}_{\;\,\ph}=G^{\theta}_{\;\,\theta}
&=& -\frac{1}{h^2}+4 (p - q) \,.
\label{Grrin}
\eea
These limit forms are shown in figures \ref{figpGtt} -- \ref{figpGtr} by the black dashed curves. Note that $\Gtata$ and $\Grr$ are discontinuous at $r=h$, but not $\Gthth$.  

Figures \ref{figfol} -- \ref{figpGtr} suggest strongly that the limit $t\to 0$ of $y(r,t)$, $g_{rr}(r,t)$ and $\Gmunu(r,t)$ in the regular co\"{o}rdinate transformation is given by the $t=0$ results of the singular transformation; we assume this to be true.

\section{Black hole features}
\label{secbhfeat}

In the previous sections we studied a regular co\"{o}rdinate transformation  and a singular one, which led to the same results in the limit $t\to 0$ for the Einstein tensor and for 
the non-temporal components of the metric. Furthermore,  the singular transformation could be tuned such that: 
\bea
g_{tt}(r,0) &=& 1/g_{rr}(r,0)\,,
\label{gttrr}\\
1/g_{rr} &=& F(r)\,,\quad\;\;\;\, h<r<\rb\,,
\label{gttsing}
\\
&=& 1-r^2/h^2\,,\quad 0<r<h\,,
\label{gttinsing}\\
g_{tr}(r,0) &=& 0\,
\eea
(Note $\rb\approx 2h$).
This metric describes at $t=0$ a black-hole with horizon radius $r=h$ and surface gravities $\kp_{\pm}$:

\bea
2\,\kp_+ &=&\lim_{r\to h^+}\partial_r g_{tt}(r,0)=F'(h)
=4 h(p-q)\,,
\nonumber\\
2\,\kp_- &=&\lim_{r\to h^-}\partial_r g_{tt}(r,0)
= - \frac{2}{h}\,.
\label{kppm}
\eea
The interior region $0 < r < h$ is like an EdS space.

For a Euclidean Schwarzschild black hole with horizon radius $r_{\rm S}$
and $r>r_{\rm S}$, $g_{tt}(r) = 1- r_{\rm S}/r$\,, the surface gravity
$\kp_{\rm S}=1/(2 r_{\rm S})$ \cite{Gibbons:1976ue}. Our choice for the integration constant $c_G$ was made by comparison of volumes as described in section \ref{seccG}, but it is interesting to see what choice of $c_G$ follows from matching the horizon radius $h=r_{\rm S}$ with the surface gravity $\kp_+=\kp_{\rm S}$: this gives
\be
c_G =
\frac{1}{2 p_0 \sqrt{p - q}}\equiv c_{G{\rm S}}\,.
\label{cGS}
\ee
With the fitted parameter values (\ref{fitrat}) $c_{G{\rm S}}=5.3$ and $h\simeq 12$, for $\kp_2=1.240$.

\begin{figure}
\includegraphics[width=7.9cm]{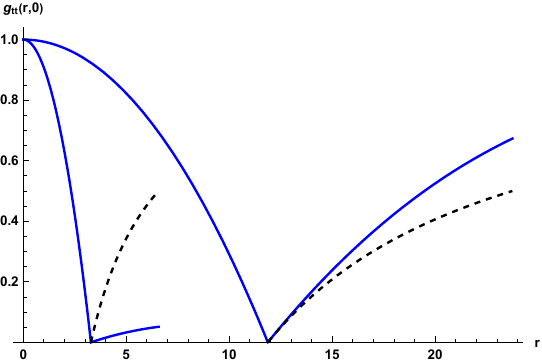} 
\caption{Metric component $g_{tt}(r,0)$ for $c_G=1.46$ ($h=3.3$) and $c_G=c_{G{\rm S}}=5.3$ ($h=11.9$), $\kp_2=1.240$. Corresponding Schwarzschild metrics are also shown (black, dashed).
}
\label{figgttS1240}
\end{figure}

Figure \ref{figgttS1240} shows $g_{tt}(r,0)$ for two values of $c_G$, together with the Schwarzschild metric (dashed). The second derivative of $g_{tt}(r,0)$ at the horizon is independent of $c_G$ and given by
\be
F^{\prime\prime}(h)=-24\, q_1\,.
\ee
Its negative sign ($-q_1 < 0$) matches the concavity of $g_{tt}$ in $r>h$, which its shares with the Schwarzschild form. 
(With the fit function $f_{\rm ch}(\et)$ the $g_{tt}$ curve has the wrong convex behavior.)

\begin{figure}
\includegraphics[width=8cm]{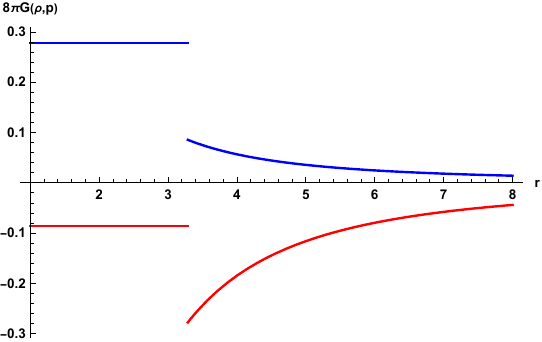}
\caption{Energy density $8\pi\GN\, \rh$ (blue, positive) and pressure $8\pi\GN\, p_r$ (red, negative) for $c_G=1.46$ ($h=3.3$).
}
\label{figpE1240}
\end{figure}

\begin{figure}
\includegraphics[width=8cm]{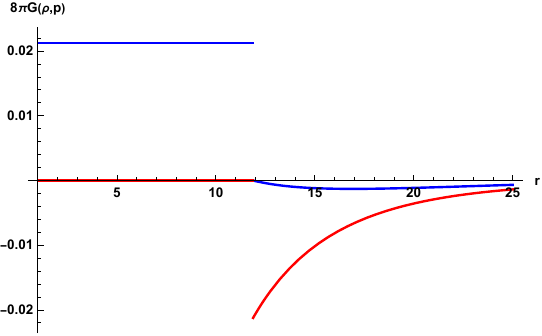}
\caption{As in figure \ref{figpE1240} for $c_G=c_{G{\rm S}}$ in (\ref{cGS}), $c_G=5.3$ ($h=12$). In this case the pressure vanishes in the interior $r<h$.
}
\label{figpE1}
\end{figure}

\begin{figure}
\includegraphics[width=8cm]{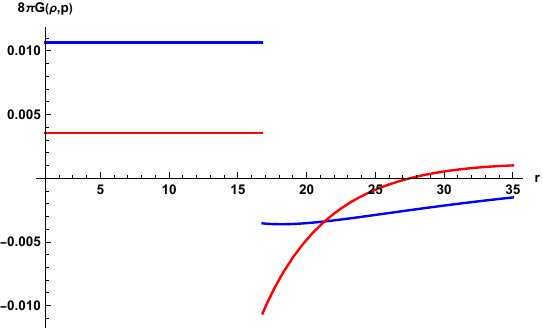}
\caption{
Energy density (blue) and pressure (red)
as in figure \ref{figpE1240} for $c_G=\cGR=7.5$ ($h=17$) in (\ref{cGR0}).
}
\label{figpER1}
\end{figure}

Assuming the Einstein equations to be valid in the effective sense, the condensate energy and pressure are given by the Einstein tensor, as in (\ref{EinsteinT}).
When $t\to 0$ it becomes diagonal with energy density and pressures given by
\be
8\pi \GN\{\rh\,,\,p_r\,,\,p_\theta\,,\,p_\ph\}=
\{- G^t_{\;\,t}\,, G^r_{\;\, r}\,,G^\theta_{\;\,\theta}\,,G^\ph_{\;\,\ph}\}\,.
\label{GET}
\ee
The interior energy-density $\rh$ is positive and the pressure $p$ is negative for $c_G < \cGS$, as shown in figure \ref{figpE1240}. Figure \ref{figpE1} shows the case $c_G=\cGS$, for which the interior pressure vanishes (note the difference in scale). For the case $c_G=\cGR$  in (\ref{cGR0}) the interior pressure is positive, as shown in figure \ref{figpER1}. In this case the curvature $R=-\sum_\mu G^\mu_{\;\,\mu}(r,0)$ vanishes in the interior and is continuous at $r=h$.

Discontinuities at $r=h$ may be expected from the discontinuous first derivative of $g_{tt}(r,0)$ at $r=h$, as evident in figure \ref{figgttS1240} and  the different surface gravities $\kp_{\pm}$ in (\ref{kppm}). The Einstein tensor contains two derivatives of the metric and when these are both spatial a Dirac delta distribution may also be present. Using the Heaviside step function $\theta(x)$, $d\theta(x)/dx=\dl(x)$, $x\,\dl(x)=0$,  and the expressions in problems 14.13 or 14.16 of \cite{Misner:1973prb}, equations (\ref{gttrr}) -- (\ref{kppm}) lead to a contribution
to the transverse pressure $p_\theta=p_\ph$,
\bea
\Delta p_\ph &=& \frac{1}{8\pi G_{\rm N}}\,
\partial_r^2 \left[(\kp_+ \theta(r-h) + \kp_-\theta(h-r))\right]
\nonumber\\
&=& \frac{\Delta\kp}{8\pi \GN}\, \dl(r-h)\,,\quad \Delta\kp=\kp_+ - \kp_- >0\,.
\label{Dirac}
\eea
On the other hand, one would expect the curvature invariant $R(r)$ to be continuous at the horizon; $R$ is equal to minus the trace of the Einstein tensor, and indeed continuous at $r=h$ by (\ref{Gtt})--(\ref{Grrin}). Since $R=8\pi \GN(\rh-p_r - p_\theta - p_\ph)$, this would imply a contribution to the energy density
\be
\Delta \rho = \frac{\Delta\kp}{4\pi \GN}\, \dl(r-h)
\ee
which cancels $\Delta p_\theta +\Delta p_\ph$ in $R$.
Integrated over the horizon this can be interpreted as a boundary layer surface tension \cite{Mazur:2015kia}.

However, if so, then one expects some regulated version of the Dirac distribution blowing up at small diminishing $t$. 
But there is no sign of this in figure \ref{figpGthth}. Indeed, as shown in appendices \ref{appET} and  \ref{appEH}, such developing singular behavior is canceled exactly by a similar contribution with two time derivatives of the metric. After all cancelations are taken into account the Einstein tensor can be expressed in a form that depends explicitly only on $r$, $\gf$, $F$ and its first derivative $F'$, but not anymore on derivatives with respect to $r$ and $t$, cf.\ (\ref{Gttapp}) -- (\ref{Einstein}). These formulas apply equally well to section \ref{secnumerical} as here and it seems remarkable that the envelope in figure \ref{figpGthth} is described correctly by the consequences of (\ref{yut}) and (\ref{yutin}). 
Thus, \emph{no evidence emerges} as $t\to 0$ for a boundary layer surface tension as in \cite{Mazur:2015kia}.

The integrated Misner-Sharp mass at $t=0$ is by 
(\ref{GET}) and (\ref{Gttin}),
\be
\lim_{r \to h^-}4\pi \int_0^r d\bar r\, r^2 \rh= \frac{h}{2\GN}\equiv M\,.
\label{MMS}
\ee
The mass (\ref{MMS}) has the Schwarzschild form.
The binding-energy estimate made in \cite{Smit:2021lyq} was $\sqrt{\GN}\approx 8$ (for $\kp_2=1.255$).
Taking $\sqrt{\GN}=10$ gives an idea of the order of magnitude of $M$ in Planck units, for our fiducial case $c_G\simeq 1.5$, $h \simeq 3.3$: $M \sqrt{\GN} \approx 0.2$.

\section{Transforming the Hayward model}
\label{secH}

\begin{figure}
\includegraphics[width=8cm]{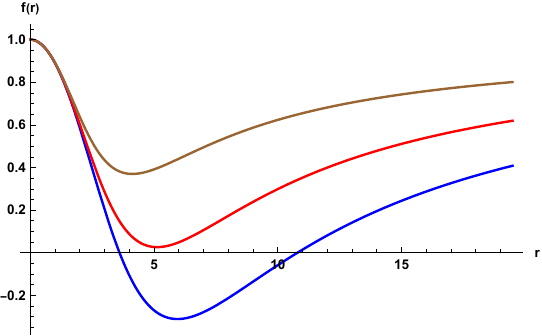} 
\caption{
Hayward function $f(r)$: top to bottom $m=0.5\,m_*$, $0.96\, m_*$, $1.5\, m_*$ and $l=3$.
}
\label{figfH}
\end{figure}

The  time-independence of the static version of Hayward's model \cite{Hayward:2005gi} allows a simple transformation to imaginary time, resulting in the Euclidean model:
\bea
ds^2 &=& f(r)\,dt^2 + \frac{1}{f(r)}\, dr^2 + r^2 \,d\Om_2^2\,,
\nonumber\\
f(r) &=& 1-\frac{2 m r^2}{2m l^2 + r^3}\,.
\label{defH}
\eea
The two parameters $m$ and $l$ are positive and have the dimension of length. For small $r$ the metric is like EdS, $f(r)\simeq 1-r^2/l^2$, and for large $r$ it is `Newton-like' $f(r)\simeq 1-2m/r$. Figure \ref{figfH} shows $f(r)$ for several values of $m$ at fixed $l$. With sufficiently large $m$ there are two real zeros at $r_+ > r_- >0$, which coincide, $r_+=r_-=r_*$, when $m$ is reduced to a critical value $m_*$. Then also $f'(r_*)=0$, which together with $f(r_*)=0$ determines
\be
m_* = \frac{3\sqrt{3}}{4}\,l\,,\quad r_*=\sqrt{3}\,l\,.
\ee

The aim here is to determine the scale factor $a(\et)$ in an $SO(4)$ invariant metric of the form (\ref{isotropic}) such that a transformation to spherical co\"{o}rdinates (\ref{spherical}) gives the metric (\ref{defH}) in the limit $t\to 0$.  For this purpose the results of the previous sections which led to $g^{rr}(r,0)=F(r)$ can be used (cf.\ (\ref{yut}) -- (\ref{uout})). Thus
\be
F(r) = f(r)\,,
\ee
and we can find $a(r)$ from the definition of $F$ in (\ref{introF}), by solving the differential equation
\be
a'(\et) =
\sqrt{f(a(\et))}\,.
\label{deqaet}
\ee
Since the Hayward model describes a regular black hole it is natural to fix the integration constant by requiring regularity at the origin,
\be
a(0)=0\,.
\label{bcaet}
\ee
To avoid a conical singularity also unit slope $a'(0)=1$ is required, which is satisfied since $f(0)=1$.

The small and large $\et$ behavior of the solution of (\ref{deqaet}), (\ref{bcaet}) are given by
\bea
a(\et) &\approx& l\,\sin(\et/l), \qquad\qquad\qquad\;\;\; a(\et)^3 \ll 2ml^2\,,
\\
&\approx& \et -m\ln(\et/m) +\mbox{const}, \quad a(\et)^3 \gg 2ml^2\,.
\label{aas}
\eea

When $m\geq m_*$,  the solution of (\ref{deqaet}), (\ref{bcaet}) has a fixed point at the first zero of $f(a)$,
\be
a(\et)\to r_-\,, \quad \et\to\infty, \quad (m\geq m_*)\,.
\ee
For $m<m_*$ but close to $m_*$ the solution can have a flat region
$a(\et)\approx  r_*$ before the asymptotic behavior (\ref{aas}) sets in.
Hence $r_*$ is similar to $h$ in the rat-model. However, the surface gravities at $r_*$ vanish,
$\kp_\pm = \lim_{r\to r_*^\pm} f'(r)/2=0$.

\begin{figure}
\includegraphics[width=8cm]{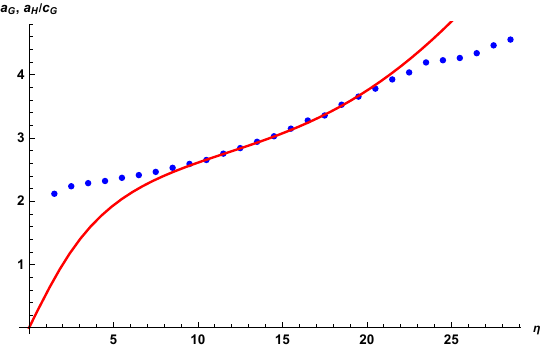} 
\caption{
Data of $a_G(\et)$ (blue) as in figure \ref{figFitsaG} matched by $(1/c_G)\, a(\et)$, with $1/c_G=0.54$ and $a(\et)$ the solution of (\ref{deqaet}), (\ref{bcaet}) with $m=0.96\, m_*$, $l=3$.
}
\label{figFitaGH}
\end{figure}

A match (by hand and eye) of $(1/c_G)\, a(\et)$ to the data in figure \ref{figFitsaG} is shown Figure \ref{figFitaGH}. Starting from the origin the curve reaches the data at much larger distances than in figure \ref{figFitsaG}. This suggests that such fits may be more appropriate in computations reaching smaller lattice spacings (in  physical units). 
At long distances $a(\et)$ approaches flat spacetime behavior according to (\ref{aas}) and has to depart from lattice data.

\section{Conclusions}
\label{secconc}

We explored the construction of an effective metric `experienced' by massless fields in the EDT formulation of quantum gravity. 
Numerical results were available from a previous study in the collapsed phase. The geometry described by the `new metric' was found to have black hole features. 

Further support of a physical interpretation of collapsed phase comes from the scaling study in appendix \ref{appEdSmetric} of the `old metric' based on the volume-distance correlator at medium distances.
The  EdS fit in this region has a curvature radius $\rnotcon$ which scales as $\rnotcon\propto N_4^{1/\ds}$ with  $\ds\approx 4.3$, or exactly 4 when incorporating a finite-volume correction (figure \ref{figpr0c}). Using $\rnotcon$ as a scaling length the metric $\acon(\et_{\rm c})$ shows good scaling (figure \ref{figrhsineerr}).\footnote{Recall $\et$ is denoted by $r$ in appendices \ref{appintroDT}, \ref{appEdSmetric}, \ref{appmasslessprop}, \ref{appcompcur}.}
The more sensitive local curvature $R_{\rm c}(\et_{\rm c})$ shows scaling but also deviations (figure \ref{figRrhsine}) which we do not understand; perhaps this observable is too ambitious for the current data, or the scaling sequence $\{N_4,\,\kp_2(N_4)\}$ is not optimal.

In our fiducial example $\{N_4,\kp_2 \} = \{32$k$, 1.24\}$, a fair match at medium distances of the volumes $V_G(\et)$ and $V(\et)$ of the new and old metrics,
\bea
V_G(\et)&=& 2\pi^2\,c_G^3 \int_0^\et d\bar \et\, a_G(\bar \et)^3\,, \\ 
V(\et)&=& N(\et)\,v_4 \,,
\label{NVGr}
\eea
made it possible to determine the integration constant $c_G$ (figure \ref{figcGet}). With this value $\simeq 1.5$,  curvatures derived from the two metrics acquired similar values at medium distances (figure \ref{figRvoldistEdS}). 

\begin{figure}
\includegraphics[width=8cm,angle=-0]{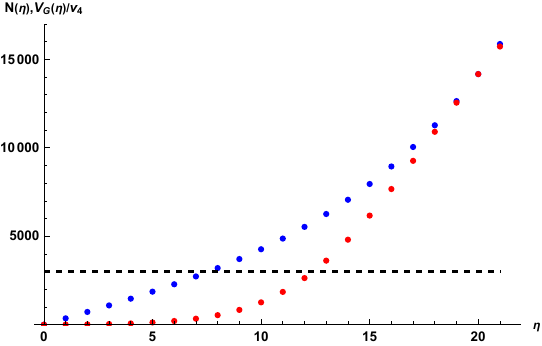} 
\caption{Comparison of $V_G(\et)/v_4$ in (\ref{NVGr}) and $N(\et)$ with $N_{\rm sing}$: top to bottom at $r=10$: $V_G(\et)/v_4$ (blue), $N_{\rm sing}$ (black dashed), $N(\et)$ (red); $N_4=32$k, $\kp_2=1.240$, $c_G=1.55$.
}
\label{figcGrsep}
\end{figure}

The quantum average blurs the position of a singular vertex relative to the origin and the way this works out differs between the two metrics. In our fiducial example there is one singular vertex with an average associated volume (the number of simplices containing it)
$N_{\rm sing} \approx 3000$ (cf.\ \cite{Catterall:1997xj}, figure 1, $\kp_0  = 2 \kp_2 = 2.48$).
When associating the singular vertex with a black hole `detected' by metric $c_G\,a_G(\et)$, $V_G(\et)$ should reach $N_{\rm sing}\,v_4$ at not large $\et$, preferably still in the short distance region. This does indeed happen for $V_G(\et)$, but not for $V(\et)$, as shown in figure \ref{figcGrsep}. In the volume-distance correlator the blurring effect seems very strong.

The blurring effect might be diminished by a `center of mass' selection as described in \cite{Ambjorn:2013eha}.
Otherwise, it seems natural to randomly choose the origin in the singular volume, perhaps averaging over it before taking the quantum average over configurations. This would partially break the discrete remnant of diffeomorphism invariance in DT. The situation appears analogous to spontaneous symmetry breaking in field theory.

The intuitive idea that the collapsed phase with its singular vertices can tell us something about black holes stimulated a change of co\"{o}rdinates that indeed showed black hole-like features, at time zero. 
We studied the effect of two types of co\"{o}rdinate transformations,
a regular one producing a diagonal metric with $g_{tt}(r,t)\neq g^{rr}(r,t)$ at all times and a singular one 
producing a metric which becomes diagonal only in the limit $t\to 0$ with $g_{tt}(r,0)=g^{rr}(r,0)$.  At time zero the component $g^{rr}(r,0)$ is the same for both transformations. As shown in figure \ref{figgttS1240} it is zero at $r=h$, as for a black hole with gravitational radius $h$. The singular transformation has a singularity at this point, which is akin to the transformation between the Schwarzschild-Droste black hole and its Kruskal-Szekeres extension
\cite{Schwarzschild:1916uq,Droste:2002,McGruder:2018bsi,Kruskal:1959vx,Szekeres:1960gm}.

The numerical study of the regular transformation showed that the left cuspy curve in figure \ref{figgttS1240} is the limiting envelope of smooth curves representing $g^{rr}(r,t)$ (figure \ref{figginv}). The time component $g_{tt}(r,t)$ differed from $g^{rr}(r,t)$,  in shape and, depending on boundary conditions specifying the implementation of the transformation, also by many orders of magnitude.
These changes in boundary conditions correspond to transformations $t\to\bar t(t)$
under which $\gf(r,t)$ and $g^{rr}(r,t)$ behave as scalar fields, but not $g_{tt}(r,t)$ which may suffer large multiplicative factors. Diagonal mixed components of the Einstein tensor, $\Gmunu(r,t)$, $\mu=\nu$,  do also transform as a scalar field under $t\to\bar t(t)$ and they approach the same zero time limit as with the singular transformation (figures \ref{figpGtt} -- \ref{figpGthth}). Remarkably, the off-diagonal $\Gtar(r,t)$ also did not suffer from the time ambiguity; it was found to vanish when $t\to 0$ and evidence was given that this holds also in the distributional sense (figure \ref{figpGtr} and appendix \ref{appEH}).

A shell-like singular distribution in the transverse pressure $\propto \Gthth(r,0)$ at $r=h$ suggests itself as a consequence of the discontinuous first derivative $\lim_{r\to h^\pm}\partial_r g_{tt}(r,0)$.
But with the regular transformation such a shell is absent due to a remarkable cancellation with a singular contribution to $\Gthth$ involving time derivatives (cf.\ appendix \ref{appEH}). One also notes that the local minimum in $g^{rr}(r,t)$ approaches $r=h$ when $t\to 0$ (figure \ref{figginv}) and the derivative at the minimum remains zero in the limit.
A gravastar-like shell \cite{Mazur:2015kia} is not supported in this study.

The co\"{o}rdinate transformations taught us that the occurrence of a horizon is closely related to the minimal derivative of $a(\et) = c_G\, a_G(\et)$, hence,  to the properties of the function $\ffit(\et)$ used to fit the discrete $a_G(\et)$ data (cf.\ (\ref{gttsing}), (\ref{introF})).
For fit functions in which the derivative $\ffit^\prime(0)$ vanishes, $h = c_G\,\ffit(0)$ is the horizon radius of a black hole, with interior region the three-sphere with radius $h$. This appears to be a quite natural possibility, which was incorporated by our choice $\ffit(\et)= f_{\rm rat}(\et)$. 

Another possibility is a derivative $\ffit^\prime(\et)$ that does not vanish at the origin and is very small at some positive $\et$. This was illustrated by applying the inverse co\"{o}rdinate transformation to the Hayward model and matching the result to the $a_G(\et)$ data at medium distances. More extensive computations are needed to determine which possibility is viable.

The method of obtaining a metric based on the propagation of massless fields may also give a new perspective of EDT's elongated phase, and the spatial collapsed-elongated boundary region (even at the transition)\footnote{Its first-order nature permits phase-coexistence, cf.\ \cite{Rindlisbacher:2015ewa} figure 8.}. 

\acknowledgements
The simulation results presented here were carried out by B.V.\ de Bakker in 1996
on the IBM SP1 and SP2 at SARA and the Parsytec PowerXplorer and CC/40 at IC3A, work supported in part by FOM  (nowadays subdivision of the Netherlands Organization for Scientific Research (NWO)).

I thank the Institute of Physics of the University of Amsterdam for its hospitality and the use of its facilities contributing to this work.

\appendix

\section{EDT and metric from a volume-distance correlator\footnote{In appendices A, B, C and D, $\et$ is denoted by $r$.}}
\label{appintroDT}

The EDT spacetimes used here are piece-wise flat manifolds consisting of equilateral 4-simplexes glued together at their 3-simplex boundaries, called `combinatorial triangulations'. The number of $i$-simplexes in these simplicial complexes is denoted by $N_i$, $i$=1,\ldots,4. In the present study their topology is $S^4$. The volume of an $i$-simplex is given by
\be
v_i=\frac{\ell^i\sqrt{i+1}}{i!\sqrt{2^i}}\,,
\ee
with $\ell$ the spacing between the lattice points. The centers of the  4-simplexes make up the dual lattice, with spacing $\ellt=\ell/\sqrt{10}$. In the following the dual lattice will be important. \emph{Unless other wise indicated we shall be using units $\ellt=1$.} 

Let $t$ denote a triangle, $t=1,\ldots,N_2$. The integral of the scalar curvature around a single triangle is given by \cite{Cheeger:1983vq,Friedberg:1984ma}
\be
\int_t d^4 x\,\sqrt{g}\,R= 2\, v_2(2\pi- o_t\,\theta)\equiv R_t\,,
\label{Rt}
\ee
where $\theta=\arccos(1/4)$ and $o_t$ is the order of the triangle, the number of 4-simplexes containing $t$.
Summing over all triangle contributions gives
\be
\intx \sqrt{g}\, R = \sum_t R_t = 2 v_2(2\pi\, N_2 - 10\,\theta\, N_4)\,.
\ee
The Einstein-Hilbert action $S_{\rm EH}$ with bare parameters $G_{{\rm N}0}$ and $\Lm_0$ becomes
\be
\frac{1}{16\pi G_{{\rm N}0}}\intx \sqrt{g}\,(2\Lm_0 - R) = \kp_4 N_4 - \kp_2 N_2\,,
\ee
where
\be
\kp_2 = \frac{v_2}{4\, G_{{\rm N}0}}\,,\quad
\kp_4 = \frac{\Lm_0 v_4 +10\theta v_2}{8\pi G_{{\rm N}0}}\,.
\ee
The partition functions are sums over triangulations $\mathcal{T}(N_4)$ at fixed topology,
\bea
Z&=&\int
Dg\, e^{-S_{\rm EH}}
\to Z(\kp_2,\kp_4)\,,
\\
Z(\kp_2,\kp_4) &=&
\sum_{N_4} e^{-\kp_4 N_4} Z(\kp_2,N_4)\,,
\label{partition1}\\
Z(\kp_2,N_4)&=& \sum_{{\cal T}(N_4)}\frac{1}{C(\mathcal{T})}\,  e^{\kp_2 N_2}\,,
\label{partition2}
\\
Z(\kp_2,\bt,N_4)&=& \sum_{{\cal T}(N_4)}\frac{ M(\mathcal{T},\bt)}{C(\mathcal{T})}\, e^{\kp_2 N_2}\,.
\label{partition3}
\eea
Here $C(\mathcal{T})$ is the order of the symmetry group (authomorphisms) that $\mathcal{T}$ may have, which is relevant for strong coupling expansions on small lattices \cite{Bilke:1998cq,Bilke:1997sc}. In (\ref{partition3}) a `measure term' \cite{Bilke:1998vj} is inserted which introduces a third coupling parameter $\bt$,
\be
 M(\mathcal{T},\bt) = \prod_t o_t^\bt = \exp\left[\bt \sum_t \ln \left(\frac{2\pi}{\theta} - \frac{R_t}{2\theta v_2}\right)\right],
 \ee
 where we used (\ref{Rt}).
It depends logarithmically on the curvature but the product form shows that is still local.

We worked with the fixed-volume, `canonical', partition function (\ref{partition2}), i.e.\ $\bt=0$.
Then the quantum average of observables $O$ is
\be
\langle O\rangle = \frac{1}{Z(\kp_2,N_4)}\sum_{{\cal T}(N_4)} e^{\kp_2 N_2}\,
O({\cal T})\,.
\label{Oav}
\ee
In particular, average scalar field propagators on $\mathcal{T}$ were computed by inverting a lattice version of the Laplace-Beltrami operator as in \cite{deBakker:1996qf}. With $S^4$ topology this operator has an exact zero mode, the constant mode. Computations of propagators in four dimensions where pioneered in \cite{Agishtein:1991cv}.

We call the average number of elementary 4-simplexes at dual lattice distance $r$ from an arbitrary origin the volume-distance correlator, $n(r)$. The average number of simplexes within distance $r$ is
\be
N(r)=N(0)+\sum_{\bar r=1}^r n(\bar r)\,,\quad N(0)=1\,,
\label{defNr}
\ee
in terms of which
\be
n(r)=N(r)-N(r-1)\equiv N'(r-1/2)\equiv \nt(r-1/2)\,.
\label{introNnt}
\ee
Here $N'$ is a discrete derivative and $\nt(r)$ is a shifted version of $n(r)$.

These observables have been compared with the volumes (\ref{contV}) of isotropic spaces in the continuum.
Typical examples are maximal symmetric spaces, which metrics were matched in \cite{Smit:2013wua} to volume-distance correlators at relatively small distances.

\begin{figure}
\includegraphics[width=8cm]{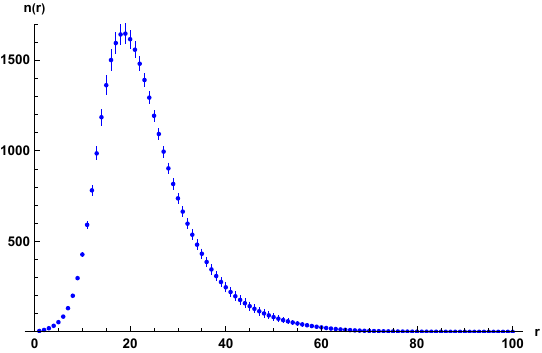} 
\caption{Volume-distance correlator $n(r)$ for $N_4=32000$, $\kp_2=1.240$; jackknife errors.}
\label{fignp240}
\end{figure}

Figure \ref{fignp240}
shows a collapsed-phase example relevant for the present work.
It can be fitted at small distances by the hyperbolic-sine form 
\be
c\,r_0^3 \sinh[(r-s)/r_0]^3\,,
\label{sinhfit}
\ee
which corresponds to a 4D hyperbolic space with constant negative curvature $-12/r_0^2$,
as follows from (\ref{Ra12}) when $a(r)=r_0\, \sinh(r/r_0)$.
The largest point in the fit domain, $2 \leq r\leq 10$, is still smaller than the inflexion point of $n(r)$ near $r=13$.
The fitted parameters are 
\be
c=0.093\,, s=-2.7\,, r_0=10.1\, \Rightarrow \lm=0.324\,, \rnotcon=3.28 
\label{csr0240}
\ee
($\lm$ and $\rnotcon$ are defined in (\ref{introlm}) and (\ref{deflm})).

At large distances $n(r)$ can be fitted by a Gaussian form 
\be 
\exp(e_0 - e_1\, r- e_2\, r^2)\,,
\label{Gaussfit}
\ee
which represents branched-polymer behavior at large $r$ \cite{Ambjorn:1995dj,Smit:2013wua}.
The fitted parameters are
\be
e_0=9.05\,,\quad e_1=0.0635\,,\quad e_2=0.000609\,.
\ee
The transition region to branched polymer behavior appears to be broad. For definiteness, the smallest distance in the fit domain was chosen to be at the right inflexion point of $n(r)$, $r=25$.
The volume of the region $r>25$ is a substantial fraction of the total volume:
$(N_4-N(25))/N_4 = 0.34$.

\begin{figure}
\includegraphics[width=8cm]{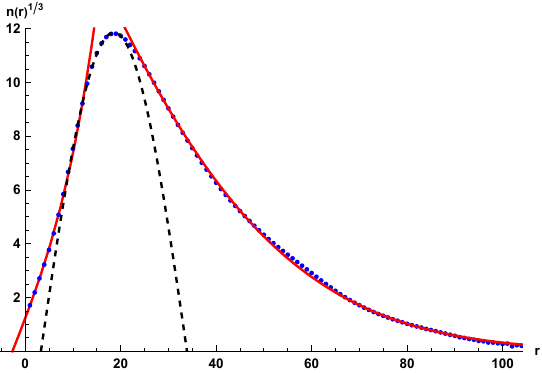} 
\caption{Central values $n(r)^{1/3}$ of the $n(r)$ data in figure \ref{fignp240} (blue dots);
$c^{1/3}\, r_0\, \sinh[(r-s)/r_0]$ from the fit to $n(r)$ by (\ref{sinhfit}) in the region $2\leq r\leq 10$
(red curve on the left flank) and $\exp[(1/3)(e_0 - e_1\, r- e_2\, r^2)]$ from the fit to $n(r)$ by (\ref{Gaussfit}) in the region $25\leq r\leq 110$.
The dashed-black curve represents the cubed-cosine fit (\ref{S4fit})$\cup$(\ref{sinecosine}) in appendix \ref{appEdSmetric}, shown for $n(r)$ in figure \ref{figfitnrcc240}. }
\label{fignpthrd240}
\end{figure}

Besides fitting $n(r)$ with constant-curvature forms, detailed scale factors $a(r)$ were also studied in \cite{Smit:2013wua} through interpolation of $n(r)^{1/3}$. Figure \ref{fignpthrd240} shows $n(r)^{1/3}$ together with the above fits to the left and right flanks of $n(r)$. 
The fit near the maximum is discussed in appendix \ref{appEdSmetric}.

In the following $\nt(r)$ is used instead of $n(r)$ as it avoids having to discuss some discretization effects later in (\ref{deflmr}).
For convenience an interpolation of $\nt(r)^{1/3}$ is denoted by $\nt_{1/3}(r)$. The lattice scale factor for a metric of the form (\ref{isotropic}) is proposed to be given by
\be
a(r-\st) =c^{-1/3}\,\nt_{1/3}(r)\,,
\label{ars}
\ee
where $c$ and $\st$ are fit parameters. For these we can use the values (\ref{csr0240}) of the short distance fit with 
fit function (\ref{sinhfit}), or its `tilde version'. Also used was a more `agnostic' version that did not involve a fit function, as follows:
linear interpolation of $\nt_{1/3}$ and extrapolation into the region $r<1/2$ results in a zero at $r=\tilde s$, $\nt_{1/3}(\tilde s)=0$, and $c$ follows from requiring
$a(r-\st)$ to vanish at $r=\st$ with unit slope,
\be
a(0)=0\,, \quad a'(0)=1\,.
\label{conda}
\ee
These requirements ensure that the space with scale factor $a(r)$ extended to $\st$ has no conical singularity.
Typically $\st$ is negative with magnitude a few $\ellt$, and for the case in figure \ref{fignp240} the parameters come out with the value of $s$ shifted by about $-1/2$ compared to (\ref{csr0240}), $\st=-3.1$, $c=0.11$\,. Subsequently the discrete values of $\nt_{1/3}(r)$ at half-integer values are interpolated by a polynomial of sufficiently high order.

\begin{figure}
\includegraphics[width=8cm]{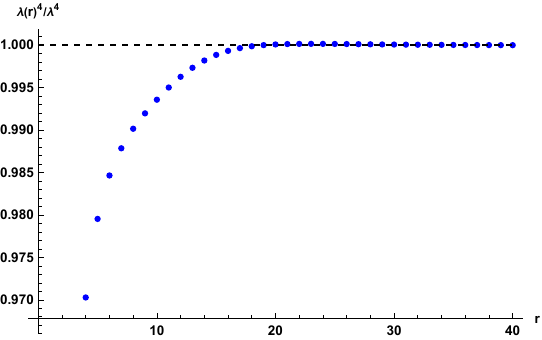} 
\caption{The ratio $\lm(r)^4/\lm^4$ in (\ref{deflmr}); $N_4=32$ k, $\kp_2=1.240$.
}
\label{figlm}
\end{figure}

To enable using continuum formulas, such as $\int d\Om_3=2\pi^2$ when calculating a volume as in (\ref{contV}), $a$ and $r$ were scaled in \cite{Smit:2013wua} by a factor $\lm$. The resulting quantities were dubbed `continuum' and indicated by a subscript `c',
\be
\rcon=\lm\, (r-s)\,, \quad \acon(\rcon) = \lm\, a(r-s)\,, \quad \rnotcon=\lm\, r_0\,,
\label{introlm}
\ee
The factor $\lm$ was defined
as
\be
\lm^4=\frac{c\, v_4}{2\pi^2}\,,
\label{deflm}
\ee
which gives $\lm=0.34$ for the case in figure \ref{fignp240}.
An $r$-dependent $\lm(r)$ can be defined by comparing the continuum volume $\Vcon(\rcon)$ with the lattice $N(r) v_4$:
\bea
\Vcon(\rcon)&=& 2\pi^2\int_0^{\rcon} d\bar\rcon\, \acon(\bar\rcon)^3
= 2\pi^2 \frac{\lm(r)^4}{c}\,\int_{\st}^{r} d\bar r\, \nt_{1/3}(\bar r)^3
\nonumber\\
&\equiv& N(r) v_4\,\Rightarrow
\frac{\lm(r)^4}{\lm^4} = \frac{N(r)}{\int_{\st}^{r} d\bar r\,
\nt_{1/3}(\bar r)^3}\,.
\label{deflmr}
\eea
The ratio $\lm(r)^4/\lm^4$ is plotted in figure \ref{figlm}; it is close to unity for
$r \gtrsim 5$.
Employing $n_{1/3}(r)$ in stead of $\nt_{1/3}(r)$ would produce a $\lm(r)$ that approaches $\lm$ at much larger distances.

\begin{figure}
\includegraphics[width=8cm,angle=-0]{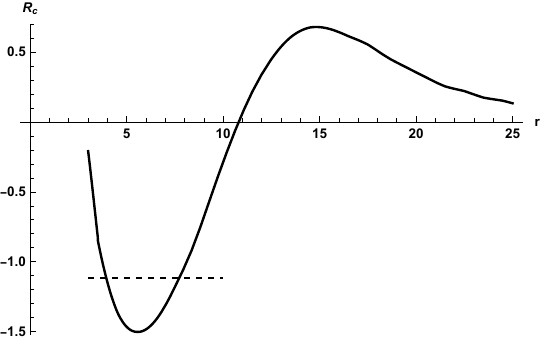} 
\caption{Curvature $R_{\rm c}$ versus $r$ from $\acon=\lm\, c^{-1/3}\,\nt_{1/3}$ with $c$ and $\lm$ from the `agnostic' values below (\ref{conda}), and $-12/\rnotcon^2$ from the short-distance fit (\ref{csr0240}).
$N_4=32$k, $\kp_2=1.240$.
}
\label{figRvoldist}
\end{figure}

The resulting scale factor $\acon(\rcon)$ can be used to compute the curvature using (\ref{Ra12}), in which we should interpret $a(r)$ as $\acon(\rcon)$.
Judging results in plots it is convenient to depict $R_{\rm c}=R/\lm^2$ as a function of the original $r$;
$R_{\rm c}=R(r-\st)/\lm^2$ is plotted in figure \ref{figRvoldist} (large wiggly black curve).
It shows negative curvature around $r=6$ and a change to positive curvature already at $r=11$.
The horizontal black dashed line in the region $r\leq 10$ is $-12/\rnotcon^2$ with curvature radius from the short-distance fit with the hyperbolic-sine function, cf.\ (\ref{sinhfit}), (\ref{csr0240}).

\section{EdS fit and scaling}
\label{appEdSmetric}

In this appendix, inspired by the results in  \cite{Laiho:2016nlp}, we concentrate on the medium-distance region.

We interpolate $n(r)^{1/3}$ and denote the result by $n_{1/3}(r)$. The positions of the maximum $\rmm$ and inflexion points $\rms^\pm$, defined by $n_{1/3}^\prime(\rmm)=0$ and $n_{1/3}^{\prime\prime}(\rms^\pm)=0$, are helpful anchors for analysis.
The medium distance region is defined as $\rms^- < r < \rmm$. Alternatively $\nt(r-1/2)=n(r)$ will be used and its results called the \emph{tilde version} in the following.\footnote{Statistical errors on our $n(r)$ data tend to make $\rms^+$ somewhat shaky but $\rmm$ and $\rms^-$ are sufficiently well determined for our purpose.}

\begin{figure}[t]
\includegraphics[width=8cm,angle=-0]{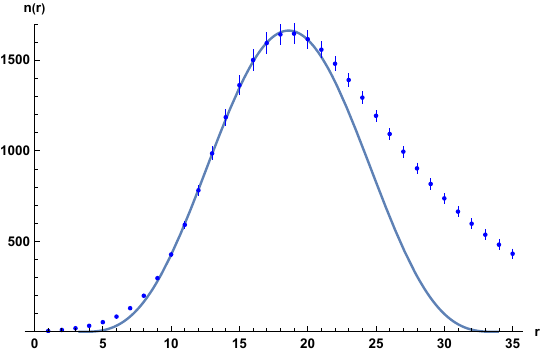} 
\caption{Plot of $n(r)$ ($\kp_2=1.24$, $N_4=32$k, $\rmm=18.6$),  fitted in the domain $r\in\{10,11,\ldots,20\}$ by $c \left[r_0 \cos((r-\rmm)/r_0)\right]^3$, $c=1.80$, $r_0=9.74$ ($s=3.30$, $\lm=0.679$).
}
\label{figfitnrcc240}
\end{figure}

In the medium-distance region we have fitted $n(r)$ by the $S^4$ fit function
\be
c\, f(r-s; r_0)=
c\, r_0^3 \,\sin[(r-s)/r_0]^3\,,
\label{S4fit}
\ee
which is derived from the EdS scale factor in the continuum,  $\acon(\rcon)=\rnotcon\,\sin(\rcon/\rnotcon)$.

Let us first summarise the construction of the continuum candidate of a metric scale factor $\acon(\rcon)$ based on the volume-distance correlator \cite{Smit:2013wua}:
\bea
n(r) &\simeq& c\,f(r-s; r_0)\,,\quad r\in\,\mbox{[fit-domain]}\,,
\label{fitnr}
\\
a(r-s)&=& c^{-1/3}\, n_{1/3}(r) \,,
\label{nca}
\\
\acon(\rcon) &=& \lm\, a(r-s)\,, \, \rcon=\lm\,(r-s)\,, \, \rnotcon = \lm\, r_0\,,
\label{aca}
\\
V_{\rm c}(\rcon)&=&2\pi^2\int_0^{\rcon} d\bar\rcon\, \acon(\bar\rcon)^3\,, \; V(r)=N(r)\, v_4\,,
\label{Vac}
\\
\Delta V_{\rm c}(\rcon) &=& \Delta V(r)\,,\quad V_{\rm c}^\prime(\rcon)=\frac{1}{\lm}\,V^\prime(r) \,,
\label{VcV}
\\
V_{\rm c}^\prime(\rcon) &=&
2\pi^2 \acon(\rcon)^3 =
\frac{v_4}{\lm}\, N^\prime(r)= \frac{v_4}{\lm}\, n(r)\,,
\label{Vcn}
\\
\lm^4 &=&\frac{c\,v_4}{2\pi^2}\,.
\label{deflm2}
\eea
In the first line (\ref{fitnr}) $f$ is a fit function, chosen to fit $n(r)$ in some domain $r\in\{\rmin,\ldots,\rmax\}$ with fit parameters $c$ and $r_0$, and possibly others such as $s$ ($f$ is assumed to vanish at $r=s$). The lattice version of the scale factor is defined in (\ref{nca}), in which $n(r)$ is supposed to be interpolated. Writing $a(r-s)$ instead of $a(r)$ is a convention.
In the third line (\ref{aca}), lengths in the lattice expressions are scaled by a factor $\lm$ to get a scale-factor that satisfies the volume `condition' $(\ref{VcV})$ which incorporates the integral over angles, $\int d\Om_3=2\pi^2$; $\Delta$ indicates the change in the fit domain, e.g.\ $\Delta V(r) = [N(r)-N(\rmin)]\,v_4$.
The second equality in (\ref{VcV}) used $d/d\rcon=(1/\lm)\,d/dr$. The relation $N^\prime(r)=n(r)$ in (\ref{Vcn}) is a choice
(in the tilde version it is a consequence of a linear interpolation of $N(r)$).
Substitution of (\ref{nca}) and (\ref{aca}) in (\ref{Vcn}) leads to the identification of $\lm$ in (\ref{deflm2}). It assumes that the implicit integral over the interpolated $n(r)$, or its tilde version, is close to $N(r)\,v_4$ for the volume `condition' $(\ref{VcV})$ to be satisfied accurately.
In any case $\acon(\rcon)$ is defined by $\lm$ in (\ref{deflm2}) together with (\ref{fitnr}) -- (\ref{aca}). It can be used to calculate other observables such as the curvature $R_{\rm c}$ using (\ref{Ra12}) with $a(r)\to \acon(\rcon)$. 

\begin{figure}
\includegraphics[width=8cm,angle=-0]{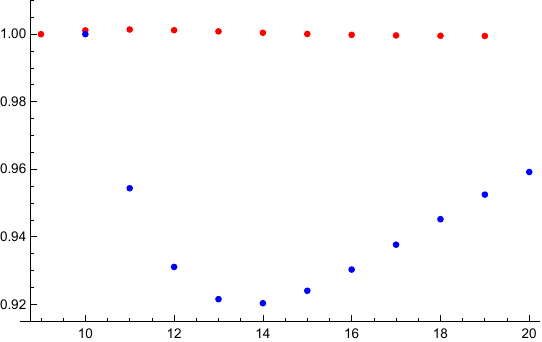} 
\caption{The ratio (\ref{ratiopcompc240}) (lower dots, blue) and its `tilde-version' (upper, red) in their respective fit domains. $\{N_4,\kp_2\}=\{32{\rm k},1.24\}$.
}
\label{figpcompc240}
\end{figure}

\begin{figure}[t]
\includegraphics[width=7cm,angle=-0]{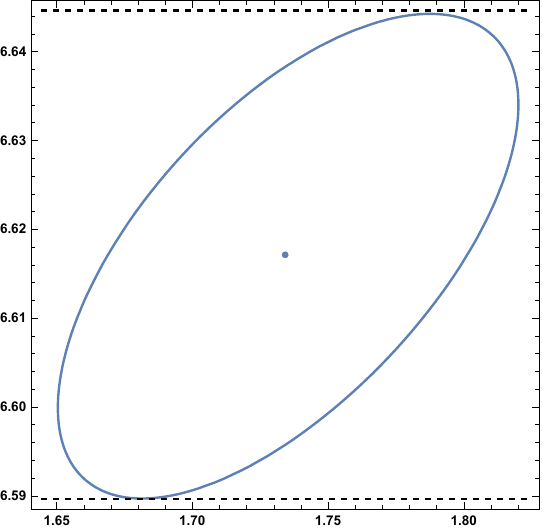} 
\caption{Contour plot of $\Delta \ch^2\equiv\ch^2-\ch^2_{\rm min}=2.30$ in the $(c,\lm(c) r_0)$ plane; $\lm(c)\equiv (v_4 c/(2\pi^2))^{1/4}$.
Dashed lines indicate error bounds $\pm 0.028$ on $\rnotcon$. The central point is at $\Delta\ch^2=0$:
$c = 1.73$, $r_0=9.84$, $\rnotcon = \lm \,  r_0 = 6.62$. The minimal $\ch^2_{\rm min} = 2.15$.
`Tilde version'. $\{N_4,\kp_2\}=\{32{\rm k},1.24\}$.
}
\label{figpcontr0r0c240t}
\end{figure}

Having determined $\rmm$ from the maximum in the interpolated $n(r)$ it is possible to avoid fits with three parameters, $\{r_0, c, s\}$ in (\ref{S4fit}), replacing them by fits with only two, $\{r_0, c\}$:
\be
\sin\left(\frac{r-s}{r_0}\right)
= \cos\left(\frac{r-\rmm}{r_0}\right)\, \&\, s=\rmm-r_0 \pi/2\,.
\label{sinecosine}
\ee

For the example $N_4 = 32$k, $\kp_2 =1.24$,
the discrete fit domain is taken to be bounded by $\rms^-$ (= 9.97) and $\rmm +1$ (= 19.6) rounded off to integers $\rmin$ and $\rmax$: $r\in \{10, 11, \ldots, 20 \}$. For the tilde version the boundary of the fit domain comes out as  $\{\rmin, \rmax\}=\{9, 19\}$. 
The fit using the equivalent cosine form in (\ref{sinecosine}) is shown in figure \ref{figfitnrcc240} and its comparison with $n(r)^{1/3}$ in figure \ref{fignpthrd240}. Note that in (\ref{sinecosine}) $ r-\rmm = r-1/2 -\tilde \rmm$ (the difference due to interpolation effects in $n_{1/3}(r)$ and $\tilde n_{1/3}(r)$ is negligible).
The fitted parameters come out as (\emph{tilde version})
\be
c=1.73\,, \, r_0=9.84\, \Rightarrow s=2.65\,, \,\lm=0.673\,;
\label{parsEdS}
\ee
$c$ and $s$ here differ qualitatively from the ones in (\ref{csr0240}).

A check on the volume `condition' (\ref{VcV}) is given by the ratio
\be
\frac{N(\rmin)v_4 + 2\pi^2 \lm \int_{\rmin}^r d\bar r\,\acon(\bar r-s)^3}{N(r) v_4}\,
\label{ratiopcompc240}
\ee
(the integral is $\Delta V_{\rm c}(\rcon)$),
which, similar to (\ref{deflmr}) should be close to 1 in the fit domain $\{\rmin, \rmin+1, \ldots, \rmax\}$.
This ratio is shown in figure \ref{figpcompc240} and we see again that the `tilde version' of the above procedure
gives  more accurate results.

The parameters $c$ and $r_0$ are quite (anti-)correlated.
The correlation is reduced between $c$ and $\rnotcon=\lm\, r_0$.
Figure \ref{figpcontr0r0c240t} shows a contour corresponding to 68.3\% confidence level if statistics is Gaussian.
The relative error of $\rnotcon$ is about half that of $r_0$ obtained from a similar plot;
the absolute error is again smaller by the factor $\lm$,
\be
\rnotcon=6.618\pm0.028\,.
\label{r0c240}
\ee

\begin{figure}[t]
\includegraphics[width=8cm,angle=-0]{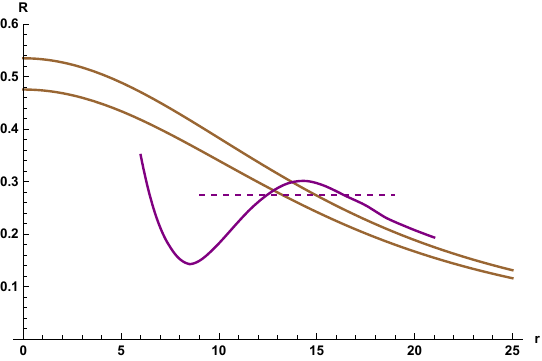} 
\caption{Curvatures versus $r$ from $\acon(\rcon)$ in (\ref{aca}) and $12/\rnotcon^2$, with $c$ and $\lm$ from the EdS fit values (\ref{parsEdS}) for $\{N_4,\kp_2\}=\{32{\rm k},1.24\}$.
Wiggly purple curve: $R_{\rm c}(\rcon)=R(r)/\lm^2$. 
Horizontal purple dashed line in $9\leq r\leq 19$: $+12/\rnotcon^2$ with $\rnotcon=6.62$ in (\ref{r0c240}). 
Also shown is curvature from the new scale factor $a(r)=c_G\,a_G(r)$ for fiducial examples of $c_G$.
The slowly varying brown curves starting near $0.5$ represent $R(r)$ with $c_G=1.46$ (upper) and 1.55 (lower).
}
\label{figRvoldistEdS}
\end{figure}

The constant curvature $+12/\rnotcon^2$ and the local curvature $R_{\rm c}=R/\lm^2$ are shown in figure \ref{figRvoldistEdS} (respectively by the purple-dashed line in $9\leq r\leq 19$ and the purple curve in $5<r<25$). For comparison the figure also shows the curvature following from the propagator metric $c_G\, a_G(r)$ with the fiducial value of $c_G$ determined in section \ref{seccG}.

Putting $\rnotcon=\lm\, s_0\, N_4^{1/4}$ can re-write the fit function:
\bea
n(r) &=& c\, r_0^3\,\sin^3\left(\frac{r-s}{r_0}\right) = \frac{\lm}{v_4}\,2\pi^2\, \rnotcon^3 \sin^3\left(\frac{\rcon}{\rnotcon}\right)
\nonumber
\\
&=& \frac{3}{4}\,\et_{\rm L}\, N_4\frac{1}{s_0 N_4^{1/4}}\,\sin^3\left(\frac{i}{s_0 N_4}+b\right),
\label{fitL}
\\
\et_{\rm L} &=&\frac{V_{\rm c}(S^4)}{V}\,,\;\; \Vcon(S^4)=\frac{8\pi^2}{3}\,\rnotcon^4\,,\;\; V=N_4\,v_4\,,
\nonumber
\\
i&=& r\,;\quad b=\frac{-\lm\,s}{\rnotcon}\,,\quad \lm\,s_0\, N_4^{1/4}=\rnotcon\,,
\eea
where $\Vcon(S^4)$ is the volume of a four-sphere with radius $\rnotcon$.
The notation in (\ref{fitL}) is that of equation (19) in \cite{Laiho:2016nlp} where this fit function was used at the boundary of the phase transition. (The identification cannot be perfect since  $i$ is an integer in \cite{Laiho:2016nlp}, which used degenerate triangulations, whereas here $r$ is integer and $\lm$ is real.)

Repeating the above procedure for other values of $\{N_4, \kp_2\}$, in our limited collapsed phase data base, we find similar results: cubed-cosine fits appear to be accurate in the medium distance region with well determined curvature radii $\rnotcon$.\footnote{A skeptic might question the interpretation of $r_0$ -- hence also $\rnotcon$ -- by claiming that a quadratic approximation $\cos(x) \to 1-x^2/2$, $x=(r-\rmm)/r_0$, gives equally good  medium-distance fits: this is not so, the cosine fits have smaller $\ch^2$. } For a sample in which the choice of $\kp_2$ was guided by seeking similar values of the fit-boundary ratios $\rms^-/\rmm\equiv \xms$, tilde-version results came out as
\be
\begin{array}{ccccccc}
\langle N_4\rangle &\kp_2&\xms&\rmm&s&\lm&\rnotcon\\
7906 & 1.17&0.472&13.9&0.8&0.571&4.746\pm 0.022\\
15909& 1.20&0.523&15.5&2.2&0.666&5.645\pm 0.013\\
31918&1.24&0.522&18.1&2.7&0.673&6.618\pm0.028\\
63911 &1.266&0.525&20.5&3.5&0.718&7.795\pm0.017
\end{array}
\label{EDTsequence}
\ee
Here $\Nav=1+\sum_r n(r)$ is the actual simulation average.

\begin{figure}
\includegraphics[width=8cm,angle=-0]{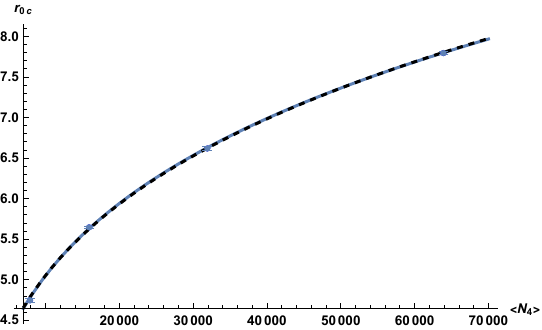} 
\caption{Scaling test of curvature radii $\rnotcon$ in the medium distance region. Data from table \ref{EDTsequence} fitted by $\rnotcon= c_{\rm s}^{\rnotcon}\, \langle N_4\rangle^{1/d_{\rm s}}$
(blue), and by $\rnotcon= c_{-1}^{\rnotcon}\, \langle N_4\rangle^{1/4} + c_{0}^{\rnotcon}$ (black-dashed);
$c_{\rm s}^{\rnotcon}=0.583$, $d_{\rm s}=4.266 \pm 0.052$, $c_{-1}^{\rnotcon}=0.466$, $c_{0}^{\rnotcon}=0.388$.  }
\label{figpr0c}
\end{figure}

Figure \ref{figpr0c} shows $\rnotcon$ versus $\langle N_4\rangle$
(blue dots with error bars). The blue curve (practically overlaid by the dashed black curve) represents a fit by the function
\be
\rnotcon= c_{\rm s}^{\rnotcon}\, \Nav^{1/d_{\rm s}}\,,
\ee
which gives $d_{\rm s} = 4.266\pm 0.052$ with a $\ch^2$ of 3.9
(the error comes from a $( c_{\rm s}^{\rnotcon},\ds)$ contour plot as in figure \ref{figpcontr0r0c240t}).
A good-looking fit ($\ch^2=5.8$) is also obtained by the first two terms of a finite-volume expansion in powers of $\Nav^{-1/4}$,
\be
\rnotcon =
c_{-1}^{\rnotcon}\,\Nav^{1/4} + c_{0}^{\rnotcon} + \cdots\,,
\label{r0cvexpansion}
\ee
as shown by the black dashed curve. The finite-volume correction $c_{0}^{\rnotcon}$ is needed here. Omitting it and trying one-parameter fits on three, or even only two, values of $N_4$ gives immediately unacceptable looking fits with at least doubled $\ch^2$ values.

\begin{figure}
\includegraphics[width=8cm,angle=-0]{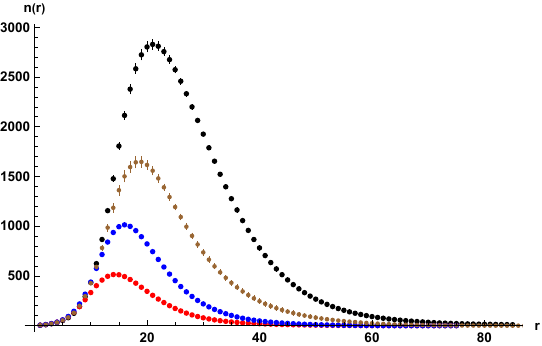} 
\caption{The four $n(r)$ of table (\ref{EDTsequence}); bottom to top at $r\approx 20$: $\kp_2=1.17$ (red), 1.20 (blue), 1.240 (brown) and 1.266 (black). }
\label{fignr}
\end{figure}

\begin{figure}
\includegraphics[width=8cm,angle=-0]{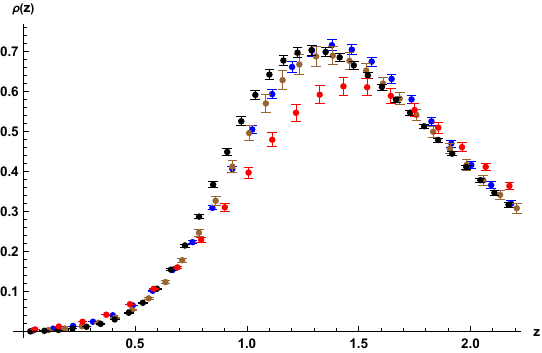} 
\caption{$\rh(z)$ with $r_*=N_4^{1/4}$ in (\ref{scalingstar}) and $N_4 \to \Nav$. Bottom to top near $z=1$: $\kp_2=1.17$, 1.240, 1.20, 1.266. Colors as in figure \ref{fignr}. }
\label{fignrsN4}
\end{figure}

The results can be used to study scaling. The $n(r)$ of table (\ref{EDTsequence}) are shown in figure \ref{fignr}.
In \cite{Smit:2013wua} we used $\rmm$ for a scaling length $r_*$ in
\be
\rh(z) = \frac{r_*}{N_4} n(z\,r_*)\,,
\label{scalingstar}
\ee
with somewhat different $\kp_2=1.21$, 1.30 and 1.260, respectively at $N_4=16$k, 32k and 64k, that were chosen to match the entire scaled $n(r)$ curves.
This $r_*= \rmm$  can be viewed as `agnostic' since it does not involve approximating $n(r)$ by some fit function.

In \cite{Laiho:2016nlp}, good scaling including all distances is obtained when using $r_* = N_4^{1/4}$ at the $\bt=0$ phase boundary, where the scaling dimension (`fractal' or `global' Hausdorff dimension) is compatible with four.
Here, using $N_4^{1/4}$ does not lead to good scaling, cf.\ figure \ref{fignrsN4}, which is no surprise given the need for a second term in the finite-volume expansion (\ref{r0cvexpansion}).

\begin{figure}
\includegraphics[width=8cm,angle=-0]{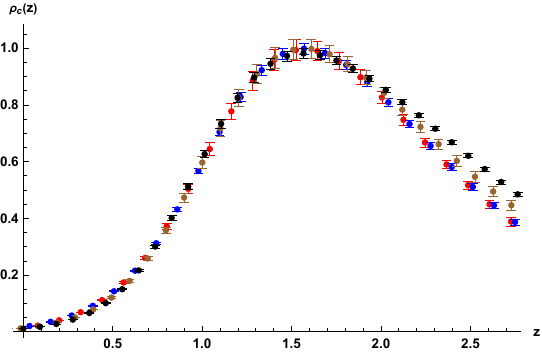} 
\caption{$\rh_{\rm c}(z)$ in (\ref{rhc}).
Top to bottom in $z\in(1, 1.1)$: $\kp_2=1.17$, 1.266, 1.240, 1.20. Colors as in figure \ref{fignr}.
}
\label{figrhsineerrn}
\end{figure}

\begin{figure}
\includegraphics[width=8cm,angle=-0]{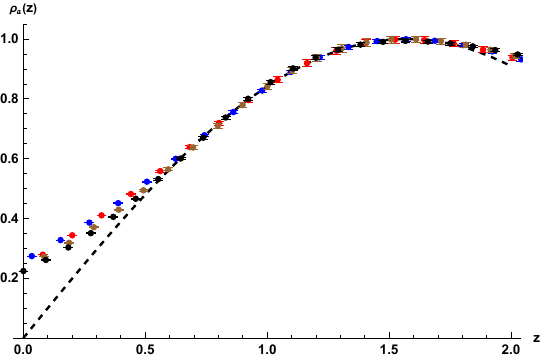} 
\caption{Scaled metric scale-factor $\rh_a(z)$ in (\ref{rhac}).
Top to bottom near $z=0.2$: $\kp_2=1.17$, 1.20, 1.240, 1.266. Colors as in figure \ref{fignr}. The black-dashed curve represents the scaled \emph{fit-functions}, $\sin(z)$. }
\label{figrhsineerr}
\end{figure}

However, $n(r)$ is not a `continuum' observable such as $\Vcon^\prime(\rcon)$, i.e.\ $2\pi^2\,\acon(\rcon)^3$. Using $\rnotcon$ 
as a scaling length, and (\ref{nca}) -- (\ref{deflm2}), in particular (\ref{aca}) for converting to the lattice $r$,
\be
\rcon = z\, \rnotcon\,, \quad r= z\, r_0 + s\,,
\ee
we arrive at the scaled observables:
\bea
\rh_{\rm c}(z)&=&  \frac{\acon( z\, \rnotcon )^3}{\rnotcon^3}=
\frac{n(z\, r_0 + s)}{c\, r_0^3}\,,
\label{rhc}
\\
\rh_a(z)&=&
\frac{\acon( z\, \rnotcon)}{\rnotcon}=
\frac{n(z\, r_0 + s)^{1/3}}{c^{1/3}\, r_0}\,,
\label{rhac}
\\
\rh_R(z)&=&\rnotcon^2\,R_{\rm c}(z\, \rnotcon)\,
\nonumber\\
&=&6\left[-\frac{\rh_a^{\prime\prime}(z)}{\rh_a(z)}- \frac{\rh_a^\prime(z)^2}{\rh_a(z)^2} + \frac{1}{\rh_a(z)^2}\right]\,.
\label{rhRc}
\eea
Note that in this scaling the argument of the volume-distance correlator involves the shift $s$: $n(z\, r_0 + s)$.

\begin{figure}
\includegraphics[width=8cm,angle=-0]{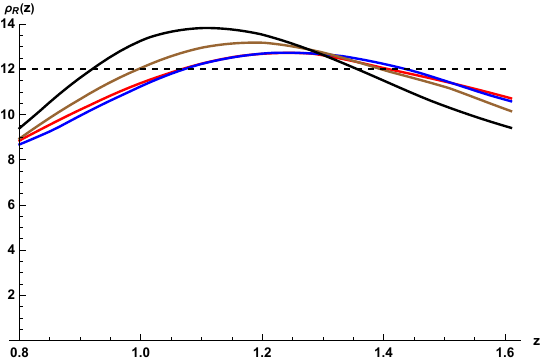} 
\caption{Scaled scalar curvature
$\rh_R(z)$ in (\ref{rhRc}) in the intersection of the scaled fit domains.
Bottom to top at $z=0.9$: $\kp_2=1.20$, 1.17, 1.240, 1.266. Colors as in figure \ref{fignr}. The constant curvature of the scaled fit functions corresponds to the dashed line.}
\label{figRrhsine}
\end{figure}

Figure \ref{figrhsineerrn} shows nice scaling for $\Vcon^\prime/(2\pi^2)=\acon^3$ at medium distances.
Figure \ref{figrhsineerr} compares the scaled $\acon$ with scaled fit functions, which are all equal to  $\sin(z)$. The scaled data appear to approach $\sin(z)$ as the volume increases.

There is no reason why the limiting $\rh_a(z)$ should become \emph{exactly} $\sin(z)$ in $0<z<\pi/2$. One would expect it does not when there is a nontrivial condensate that differs from a mere cosmological constant.

As shown in figure \ref{figRrhsine}, the scaled local curvatures $\rnotcon^2\, R_{\rm c}(\rcon)$  of the two smallest volumes match closely in the intersection of the fit domains $0.80<z< 1.61$, but those of two larger volumes differ substantially. For the case $\{N_4=32{\rm k}, \kp_2=1.24\}$ the curvature $R_{\rm c}$ is also shown in figure \ref{figRvoldistEdS} as a function of $r$ (in this case the intersection region corresponds to $11<r<19$).

\section{Massless propagators  \cite{deBakker:1996unpub}}
\label{appmasslessprop}

On a space with constant negative curvature, a massless propagator
shows exponential fall off with the distance and can thus be viewed as
having a mass.

Rewriting the equation for the massless propagator
$\square\, G(x^\mu) = 0$ using spherical symmetry and a space of constant negative curvature
with curvature radius $r_0$ we get
\begin{equation}
  \left[ \frac{d}{dr} + \frac{3 \coth(r/r_0)}{r_0} \right] \frac{d}{dr}
  G(r) = 0 \,.
\end{equation}
This can be exactly solved and using the boundary conditions
\begin{eqnarray}
G(r) &\to& 0\,,\qquad\; r \to \infty \,, \\
G(r) &\to& \frac{1}{4 \pi r^2}\,,\quad r \to 0\,,
\end{eqnarray}
and the solution is
\begin{eqnarray}
  G(r) &=& \frac{1}{4 \pi r_0^2} \left\{ 2 \frac{\exp(r/r_0) +
  \exp(-r/r_0)}{ \left( \exp(r/r_0) - \exp(-r/r_0) \right)^2 }
  \right.
  \\&&\left. - \ln
  \frac{1+\exp(-r/r_0)}{1-\exp(-r/r_0)} \right\} .
\label{GnegR}
\end{eqnarray}
Expanding this around $r \to \infty$ in $\exp(-r/r_0)$ results in
\begin{equation}
  G(r) = \frac{2}{4 \pi r_0^2} \sum_{n = 0}^{\infty} \frac{n(n+1)}{2n+1}
  \exp(-(2n+1)r/r_0) \,.
\end{equation}
We see that for large $r$ the massless propagator behaves as
\begin{equation}
  G(r) \propto \exp(-3r/r_0), \quad r \to \infty \,.
\label{GnegRas}
\end{equation}
If we define the mass $m$ from the long distance behaviour of the
propagator using $G(r) \propto \exp(-mr)$, then this mass and the
curvature radius are related by $m r_0 = 3$.

\begin{figure}
\includegraphics[width=8cm]{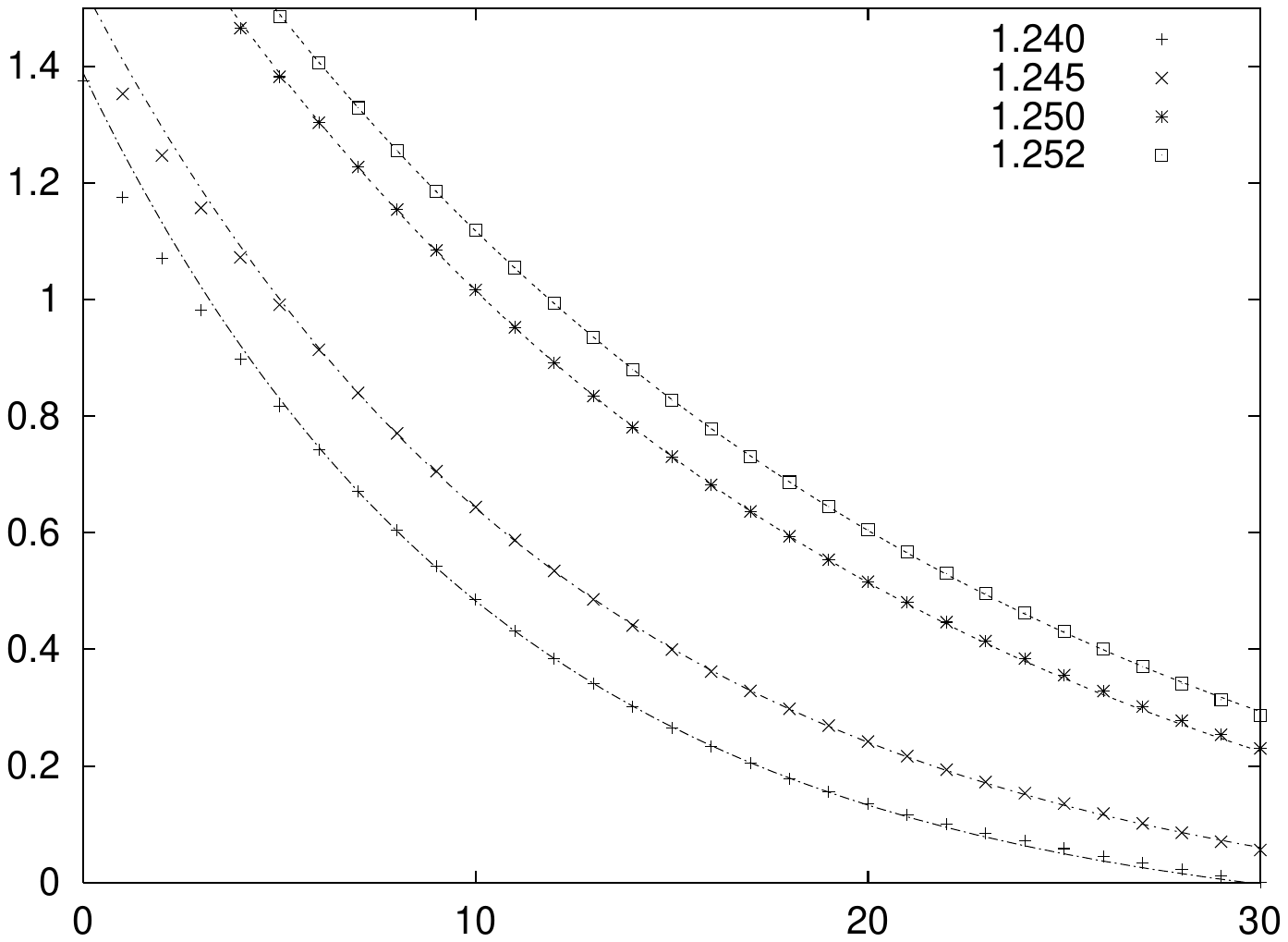}
\caption{Averaged massless lattice propagators $G(r)$ for some values of $\kappa_2$ at
$N_4 = 32000$.}
\label{propfig}
\end{figure}

On the lattice, we solved the Laplace equation $\square\, G(x^\mu) = 0$
using the algebraic multigrid routine AMG1R5.  Figure~\ref{propfig} shows the results for the
smallest volume $N_4 = 32000$.  Note that we can add an arbitrary
constant to $G(r)$.  This constant is determined by the details of the
computer code and has no physical meaning.  This also means that the
individual values of the data points have no meaning and therefore we
have left out the error bars.  This is not an obstacle to calculating
the masses and their errors.

We fitted this data to $a \exp(-mr) + c$ in the range
$6 \leq r \leq 20$.
The fits are also shown in figure~\ref{propfig}.

\begin{figure}
\includegraphics[width=8cm]{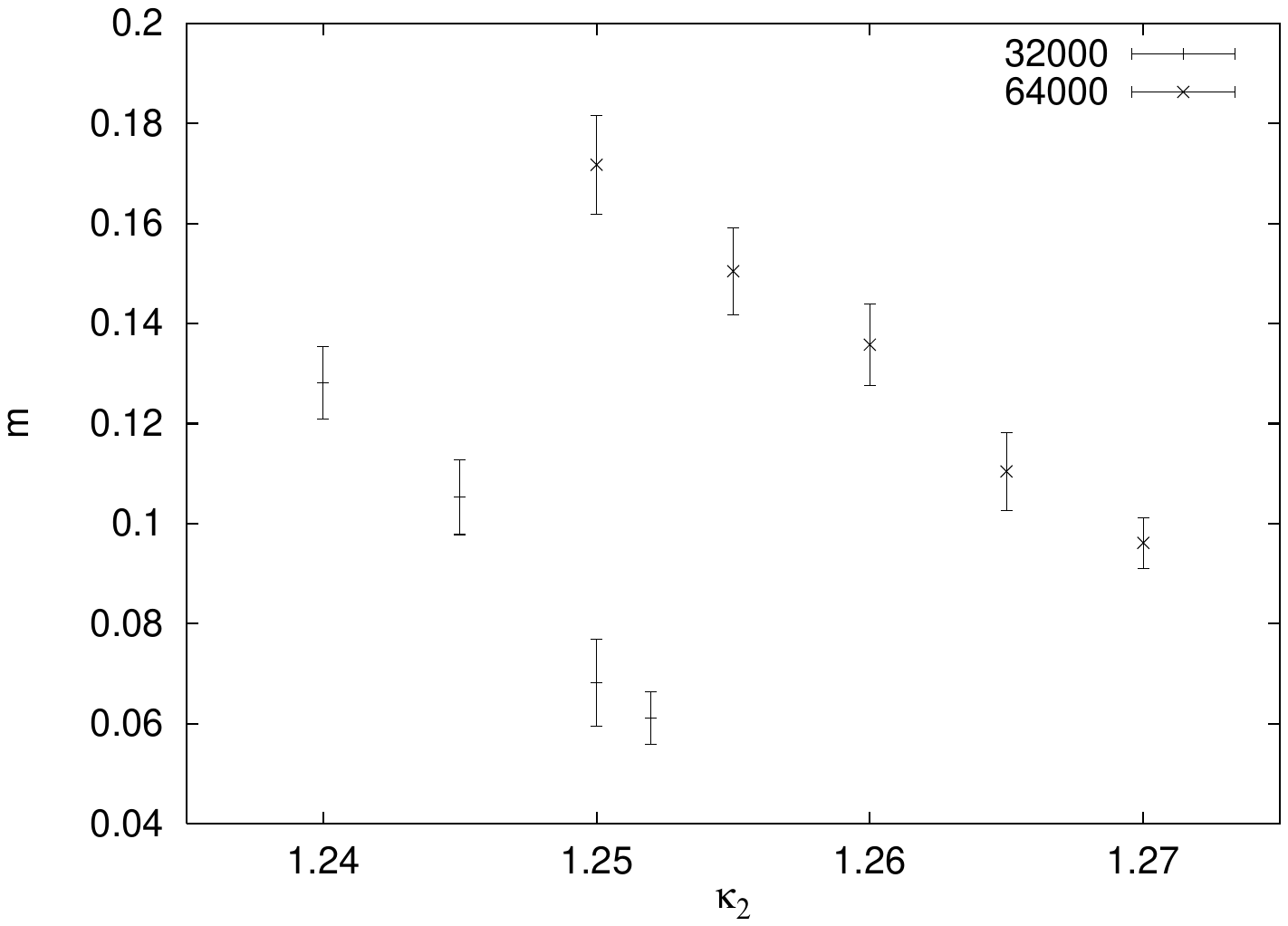}
\caption{Measured effective masses of propagators as a function of $\kappa_2$
for two values of $N_4$.}
\label{massfig}
\end{figure}

Figure~\ref{massfig} show the masses $m$ measured for various values
of $\kappa_2$ at the two volumes that were used.  The error bars were
obtained by the jackknife method, leaving out one configuration each
time.  In this way the correlation between the data of two different
origins on the same configuration does not artificially decrease the
errors.

The configurations used for these measurements were recorded each
$5000$ sweeps, where a sweep consists of $N_4$ accepted moves.  For
each configuration we used 200 different origins to calculate the
propagator from.  We could only measure these masses for a limited
number of $\kappa_2$ values, because the calculation of the propagator
takes much CPU time.

\section{Comparison of effective masses with curvature radii}
\label{appcompcur}

The original motivation of the propagator study was the negative curvature found in the volume-distance correlation at short distances. In this appendix the exponential behavior of the propagator data is analyzed further.

\begin{figure}
\includegraphics[width=8cm]{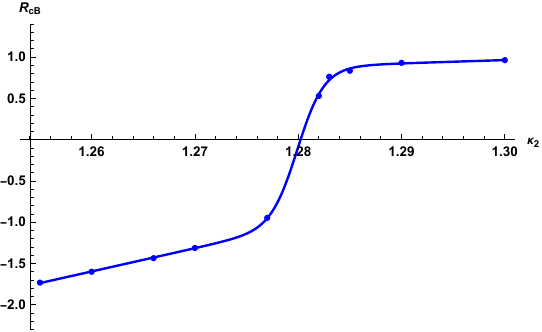} 
\caption{
Curvature $R_{\rm c}$  vs.\ $\kp_2$ obtained with the B-fit (dots) fitted by the function (8.6) in \cite{Smit:2013wua}: $N_4=64$k, data from the Erratum.
}
\label{figRcB}
\end{figure}

For a comparison with the asymptotic form (\ref{GnegRas}) that predicts $m r_0/3=1$, we use $r_0$ results recorded in table (B.2) of \cite{Smit:2013wua} for the volume $N_4=64$k. Figure \ref{figRcB} shows a (so-called `continuum curvature') $R_{\rm c}$ in both phases.\footnote{From figure 14 in \cite{Smit:2013wua} we see that for $N_4=64$k all the $\kp_2$ values in figure \ref{massfig} lay outside the curvature-susceptibility peak on the collapsed side; for $N_4=32$k this holds only for $\kp_2=1.240$ and 1.245, whereas 1.250 and 1.252 are within the peak.}

Numerical data files of the propagators and fitted parameters of $a \exp(-m r) + c$ are lost, unfortunately. However, we have retrieved values of $m$, and for $N_4=32$k also of the propagators, from the postscript files of figure \ref{propfig} and \ref{massfig}.

\begin{figure}
\includegraphics[width=8cm]{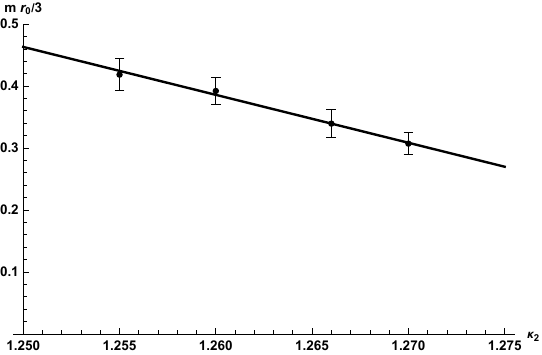} 
\caption{
The ratio $m/(3/r_0)$ vs.\ $\kp_2$ in the collapsed phase with $r_0$ values from the fit in figure \ref{figRcB}; $N_4=64$k. The line is a linear fit by $10-7.7\, \kp_2$. }
\label{figmr0}
\end{figure}

Figure \ref{figmr0} shows the ratio $m/(3/ r_0)$ vs.\ $\kp_2$ with the $r_0$ corresponding to figure \ref{figRcB} in which $N_4=64$k. (With $N_4=32$k only $\kp_2=1.240$ and 1.245 volume-distance results are available.)
The linearly varying ratio seems to point again to negative curvature. The magnitude of the ratio is much smaller than 1.
An immediate explanation is suggested by the fact that the effective-mass fits to the propagators involve larger distances (up to $r=20$) than the negative-curvature fits to the volume-distance correlator which ended at $r=12$. We shall conclude below that the discrepancy is even more fundamental.

\begin{figure}
\includegraphics[width=8cm]{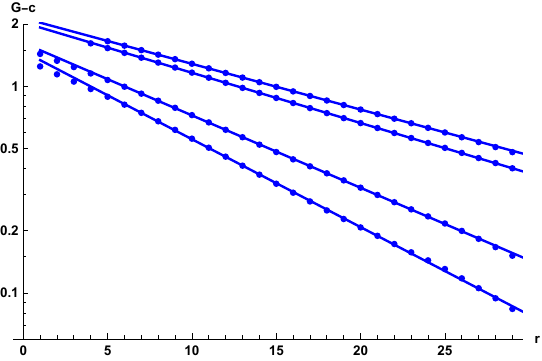} 
\caption{Propagators fitted by $c+ a\exp(-m r)$ in the domain $6\leq r\leq 24$. Data and fit-function are both subtracted by $c$; top to bottom $\kp_2= 1.252$, 1.250, 1.245, 1.240; $N_4=32$k.
}
\label{figprops}
\end{figure}

Refitting the retrieved propagator data with the fit function $a \exp(-m r) + c$ in the domain
$6\leq r\leq 24$, the fitted parameters came out as  ($N_4=32$ k):
\be
\begin{array}{cccccccccc}
\kp_2&c&a&m&\reff\equiv 3/m&
\\
1.240&-0.0726&1.48&0.098&30.7\\
1.245&-0.0816&1.62&0.081&37.2\\
1.250&-0.148&2.04&0.056&53.5\\
1.252&-0.167&2.14&0.051&58.9
\end{array}
\label{parspropfit}
\ee

The effective curvature radii $\reff$ are remarkably large.
Our refitted $m$ are 20-30 \% smaller than those in figure \ref{massfig} (the original fit domain $6\leq r\leq 20$ in section \ref{appmasslessprop} leads to even slightly smaller values and spoils the good looking fits in figure \ref{figprops} for $r>20$).
The re-fits are of the least-squares type whereas the error bars in figure \ref{massfig} represent jack-knife errors, which tend to underestimate errors.
Overlaying a print of the plot in figure \ref{propfig} by a same size print for the propagators obtained with refitted parameters gives a perfect match visually.
We ascribe the difference to small mis-representations by the postscript files together with correlations between the 
fit parameters $a$, $m$ and $c$. It does not affect the exploring nature of this work.\footnote{Another example is the fit value of $e_1$ in figure \ref{figFitaGl}, which corresponds to an $m=3 e_1=0.11$ differing 8\% from the 0.98 in (\ref{parspropfit}).}

\begin{figure}
\includegraphics[width=8cm]{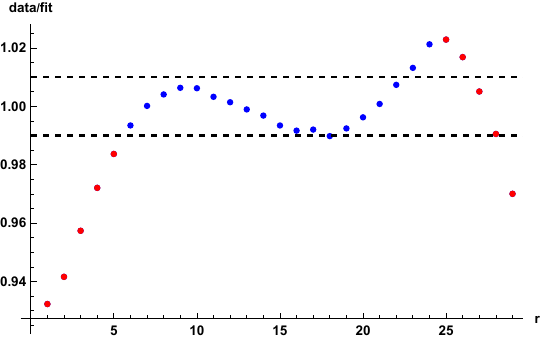} 
  \caption{
Subtracted propagator $G(r)-c$ divided by subtracted fit-function $a \exp(-m r)$, for $\kp_2=1.240$. The blue(red) points are inside(outside) the fit domain.}
\label{figdataoverfit}
\end{figure} 

Figure \ref{figprops} shows the obtained propagators shifted by $-c$ on a logarithmic scale. Exponential behavior is a fairly good approximation as can be seen quantitatively in figure \ref{figdataoverfit} which shows the ratio (data-$c$)/($a\exp(-m r))$ for $\kp_2=1.240$.
In this case the deviations from 1 are at the percent level in the fit domain; for the other $\kp_2$ such deviations are smaller by an order of magnitude. The local minimum near $r=18$  occurs where the volume-distance correlator has its maximum (cf.\ figure \ref{fignp240} in appendix \ref{appintroDT}).

Remarkably, this maximum hardly affects the nearly exponential behavior of the shifted propagators.
Surprising is also the fact that this behavior occurs all the way down to about $r=6$. It implies that using the exact form (\ref{GnegR}) of $G(r)$ in a fit function of the form $c+a\, G(r)$ {\em cannot fit the data}. This becomes also immediately clear when trying such fits. Last, but not least, one notes that $G(r)$ in (\ref{GnegR}) approaches the leading term in its asymptotic expansion to 1\% accuracy only for $r/r_0 > 2.59$; for $\kp_2=1.240$ this means $r > 80$ ! 

The goodness of the exponential fit does \emph{not} indicate hyperbolic nature of the geometry.

\section{Einstein tensor for a time-dependent diagonal metric}
\label{appET}
In equation (14.48) of \cite{Misner:1973prb}, a time-dependent rotation-invariant diagonal Lorentzian metric depending on  co{\"o}rdinates $T$, $R$, $\theta$ and $\ph$ is defined by its line element
\be
ds^2 = -e^{2\Ph}\,dT^2 + e^{2\Lm}\, dR^2 + r^2 (d\theta^2 + \sin(\theta)^2 d\ph^2)\,,
\ee
and its Einstein tensor is given in (14.48), (14.51) and (14.52) of this book. Transforming co{\"o}rdinates $T=-i\,t$ and choosing $R=r$ we obtain the Einstein tensor for the Euclidean metric.
The transformed functions of \cite{Misner:1973prb} are given by
\be
\Ph = \half\,\ln g_{tt}\,,\quad \Lm = \half\,\ln g_{rr}\,,
\ee
and
\bea
E_1 &=& e^{-2\Ph} (\ddot \Lm + \dot\Lm^2 -\dot\Lm \dot \Ph)\,,
\\
E_2&=&-e^{-2\Lm} (\Ph''+\Ph'^2-\Ph^{\prime}\Lm')\,,
\\
E &=& E_1 + E_2\,,\quad \bar E = -\frac{1}{r}\,e^{-2\Lm}\; \Ph'\,,
\\
F_{\mbox{\tiny{MTW}}}&=&\frac{1}{r^2}\,(1-e^{-2\Lm})\,,\quad \bar F_{\mbox{\tiny{MTW}}} = \frac{1}{r}\,e^{-2\Lm}\; \Lm'\,,
\\
H&=& -i \frac{1}{r}\,e^{-\Ph-\Lm}\; \dot\Lm\,,
\eea
in terms of which Euclidean mixed (up-down) components of the Einstein tensor are given by
\bea
G^t_{\;\,t} &=& - F_{\mbox{\tiny{MTW}}} - 2\bar F_{\mbox{\tiny{MTW}}}\,,\quad \Grr= -2\bar E - F_{\mbox{\tiny{MTW}}}\,,
\nonumber
\\
\Gthth &=&\Gphph = -E - \bar E - \bar F_{\mbox{\tiny{MTW}}} \,,\quad \Gtar= 2\, i H\,.
\label{EGMTW}
\eea
Note that $E_1$ contains two time derivatives of $g_{rr}$, $E_2$ two spatial derivatives of $g_{tt}$, and that their sum $E_1+E_2$ enters in $\Gthth$.
It is convenient to treat $F(a)$ introduced in (\ref{introF}) as a function of $a^2=r^2 + \gf^2$, writing
\be
F(\sqrt{r^2 + \gf^2}) = \Ft(r^2 + \gf^2)\,.
\label{introtF}
\ee
The functions $E$, \ldots, $H$ depend on derivatives of the metric components
\be
\fg \equiv g_{tt}\,,\quad \fgi \equiv \frac{1}{g_{rr}}
\ee
(cf.\ (\ref{gtt}), (\ref{grr})). Indicating their explicit dependence on variables these are given by
\bea
\fgi(r,\gf)&=& \frac{r^2 \Ft(r^2 + \gf^2) + \gf^2}{r^2 + \gf^2}\,,
\\
\fg(r,\gf,\dot\gf) &=& \frac{\dot\gf^2}{\Ft(r^2 + \gf^2)}\,\fgi(r,\gf)\,.
\eea
Since $\gf$ is a solution of $\partial_r\gf=P$,
\be
P(r,\gf)= r\gf \frac{-1+\Ft(r^2 + \gf^2)}{\gf^2 + r^2 \Ft(r^2 + \gf^2)}\,,
\ee
we can replace spatial derivatives $\partial_r\gf$ by $P$:
\bea
\frac{d}{dr}\,\fgi &=& \frac{\partial \fgi}{\partial r} + \frac{\partial \fgi}{\partial \gf}\, P\,,
\\
\frac{d}{dr}\,
\fg &=& \frac{\partial \fg}{\partial r} + \frac{\partial \fg}{\partial \gf}\, P
+ \frac{\partial \fg}{\partial \dot\gf}\, \partial_r \dot\gf\,,
\\
\partial_r\,\dot\gf
&=&\partial_r\partial_t \gf
= \partial_t \partial_r \gf =
\frac{d}{d t}\, P = \frac{\partial P}{\partial \gf}\, \dot \gf\,,
\eea
and recursively for higher derivatives. In principle this gives $E_1$, $E_2$, \ldots, $H$ as functions depending explicitly only on $r$, $\gf$, $\dot\gf$, $\ddot\gf$ ($\dddot\gf$ is not needed since double time derivatives occur only on $\Lm$ which does not contain $\dot\gf$). However the explicit dependence on  $\dot\gf$ and $\ddot \gf$ cancels out. This should happen in the diagonal components of the Einstein tensor which transform as a scalar field under time transformations $\bar t(t)$ and $\dot\gf$ and $\ddot\gf$ are not such scalars. But it happens already in the individual $E_1$, $E_2$, \ldots, $H$.

Further details of $E_1$, \ldots, $H$ are in appendix \ref{appEH}.
There is a near cancellation between $E_1$ and $E_2$, hence also in their sum $E=E_1 + E_2$ contributing to $\Gthth$. This is relevant because $E_1$ and $E_2$ separately are increasingly strongly peaked as $t\to 0$, whereas $E$ clearly reaches a finite limiting form.
The Dirac distribution $\dl(r-h)$ conjectured in (\ref{Dirac}) as a result of the double spatial derivative in $E_2$, is not present in $E$, hence also not in $\Gthth$ at $t=0$.    We were not able to prove that $\dl(r-h)$ emerges in $E_1$ and $E_2$ separately as $t\to 0$, although there is modest numerical evidence for a finite limit in the distributional sense.

The components of $\Gmunu$ listed in (\ref{EGMTW}) become:
\bea
\Gtata&=& \frac{(\Ft-1)(3 \gf^2 + r^2 \Ft)}{(\gf^2 + r^2) (\gf^2 + r^2 \Ft)}
+ \frac{2 r^2 \Ft \Ft'}{\gf^2 + r^2 \Ft}\,,
\label{Gttapp}
\\
\Grr &=&\frac{(\Ft-1)(\gf^2 + 3 r^2 \Ft)}{(\gf^2 + r^2) (\gf^2 + r^2 \Ft)}
+\frac{2 \gf^2 \Ft'}{\gf^2 + r^2 \Ft}\,,
\label{Grrapp}
\\
\Gthth &=& \Gphph =\frac{\Ft-1}{\gf^2 + r^2} + 2 \Ft'\,,
\label{Gththapp}
\\
\Gtar &=& 2 r \gf \sqrt{\Ft}\left[\frac{\Ft-1}{(\gf^2 + r^2) (\gf^2 + r^2 \Ft) }
- \frac{\Ft'}{\gf^2 + r^2 \Ft}\right]\,.
\nonumber
\\
\label{Einstein}
\eea
Here $\Ft'$ is the derivative of $\Ft$ with respect to its argument: $\Ft'(r^2 + \gf^2)=d \Ft(x)/dx|x\to r^2 + \gf^2$ (and similar for $\Ft^{\prime\prime}$ which appears in $E_1$ and $E_2$). The trace of the Einstein tensor simplifies to
\be
\Gmumu = \frac{6 (-1 + \Ft)}{\gf^2 + r^2} +  6\, \Ft^\prime \,.
\ee
Following the reasoning in section \ref{secttozero}, the limit-forms for $t\to 0$ in that section follow here easily -- without encountering singularities -- by letting $\gf\to 0$ in the exterior region, and
$\Ft\to 0$ \& $\gf\to \sqrt{h^2 - r^2}$ in the interior region. The off-diagonal component $\Gtar$ vanishes in the limit. The possibility of a remaining finite distribution at $r=h$ is investigated in appendix \ref{appEH}.

\section{Details of $E_1$, \ldots, $H$}
\label{appEH}

 \begin{figure}
\includegraphics[width=8cm]{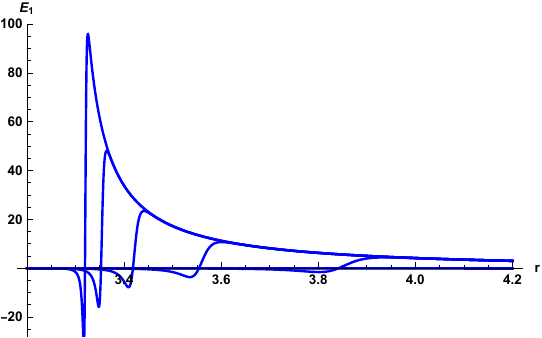} 
  \caption{Plot of $E_1(r,t)$ for $\rref=2 h$ and $t= 10^{k-24}$, $k=0,4,8,12,16$.
  }
\label{figE1}
\end{figure}

\begin{figure}
\includegraphics[width=8cm]{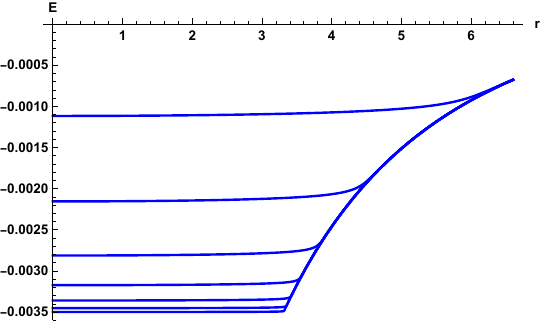} 
  \caption{Plot of $E=E_1+E_2$ for the same times as in figure \ref{figE1}.
  }
\label{figE}
\end{figure}

After the canceling-out of $\dot\gf$ and $\ddot\gf$, $E_1$ and $E_2$ are given by
\bea
E_1 &=& e_{10} + e_{11}\,\Ft'+ e_{12}\, (\Ft^{\prime})^2\,,
\nonumber\\
E_2 &=& e_{20} + e_{21}\, \Ft' + e_{22}\, (\Ft^{\prime})^2\,,
\nonumber\\
e_{10} &=& \frac{1}{(y^2 + \Ft r^2)^3}\left[r^2 (3 y^2 (-1 + \Ft) \Ft - (-1 + \Ft) \Ft^2 r^2)\right.
\nonumber\\
&&\left. +
 \Ft^{\prime\prime} r^2 (2 y^6 \Ft + 2 y^4 \Ft (1 + \Ft) r^2 + 2 y^2 \Ft^2 r^4)\right]\,,
 \nonumber\\
 e_{11} &=& \frac{1}{(y^2 + \Ft r^2)^3}\left[r^2 (\gf^4 (1 - 4 \Ft) + \Ft^2 r^4 \right.
 \nonumber\\
 &&\left.  + \gf^2 \Ft (-6 r^2 + 4 \Ft r^2))\right]\,,
 \nonumber\\
 e_{12} &=& \frac{1}{(y^2 + \Ft r^2)^3}\left[r^2 (\gf^6 + \gf^4 (1 - 3 \Ft) r^2 - 3 \gf^2 \Ft r^4)\right]\,,
 \nonumber\\
 e_{21}&=&  \frac{1}{(y^2 + \Ft r^2)^3}\left[(-\gf^6 + \gf^4 (-1 + \Ft) r^2 \right.
 \nonumber\\
 &&\left. + \gf^2 \Ft r^2 (6 r^2 - 7 \Ft r^2) - \Ft^2 r^4 (r^2 + \Ft r^2))\right]\,,
 \nonumber\\
 e_{20} &=& -e_{10}\,,\quad e_{22}=-e_{12}\,,\quad
 e_{11} + e_{21} = -1\,,
 \nonumber\\
 E &=& E_1 + E_2 = -\Ft'\,.
 \eea
 The expressions for  $\bar E$, \ldots, $H$ come out as:
\bea
\bar E &=& -\frac{(-1 + \Ft) \Ft r^2}{(\gf^2 + r^2) (\gf^2 + \Ft r^2)}
-\frac{\gf^2 \Ft^\prime}{\gf^2 + \Ft r^2} \,,
\nonumber\\
F_{\mbox{\tiny{MTW}}} &=& \frac{1 - \Ft}{\gf^2 + r^2}\,,
\nonumber\\
\bar F_{\mbox{\tiny{MTW}}} &=& -\frac{\gf^2 (-1 + \Ft)}{(\gf^2 + r^2) (\gf^2 + \Ft r^2)}
-\frac{\Ft \Ft^\prime r^2}{\gf^2 + \Ft r^2}\,,
\nonumber\\
i H &=&r \gf\sqrt{\Ft}\left[\frac{\Ft-1}{(\gf^2+r^2)(\gf^2 + \Ft r^2)} -\frac{\Ft^\prime}{\gf^2 + \Ft r^2}\right]\,.
\nonumber\\&&
\eea

For the rat-model $\Ft$ and $\Ft'$ can be written in the form (cf.\ (\ref{Frat}))
\bea
\Ft&=&c (\sqrt{x}-h) (\sqrt{x} - \bar h)^3\,,\quad x=r^2 + \gf^2\,,
\\
\Ft'&=& \frac{c (4\sqrt{x}-3 h - \bar h)(\sqrt{x}-\bar h)^2}{2\sqrt{x}}\,.
 \eea

Figure \ref{figE1} shows a plot of $E_1$ corresponding to five (blue) curves of $\gf(r,t)$ in figure \ref{figfol} with $\rref=2h$.  A similar plot for $-E_2(r,t)$ is indistinguishable to the eye, since the sum $E_1 + E_2$ is down in magnitude by a factor of about $10^4$, note the vertical scale in figure \ref{figE} which displays $E$.

The integral $I=\int_{3.2}^{4.2} dr\, E_1(r,t)$ was monitored to check wether a finite distribution (such as a Dirac function $\dl(r-h)$) develops in $E_1$ (and also in $E_2$ as follows from the cancelation) in the limit $t\to 0$. Considered as function of $\ln(t)$, $I$ is very well fitted by the form $I=\al + \bt \ln(t)$, which suggests a logarithmic divergence in the limit. But a dependence of $I$ as a function of a time $t$ introduced at $\rref=2h$, `far away' from $r=h$ involves a  co{\"o}rdinate peculiarity of this $t$ (cf.\ figure \ref{figfollog}). Testing as a function of the foliation as labeled by the value of $\gf$ at $r=h$, i.e.\ $\gf_t=\gf(h,t)$, may be a better idea. The values of $I$ are well fitted by the rational-function form
$I=(\al +\bt\, \gf_t)/(1+\gm\, \gf_t)$, which has a build-in finite limit as $\gf_t\to 0$. We take this as a mild support for a finite limit distribution $E_{1,2}$ at $r=h$.  However, since only the regular sum $E$ enters in $\Gthth$ the finiteness of $E_{1,2}$ is not of physical interest.

\begin{figure}
\includegraphics[width=8cm]{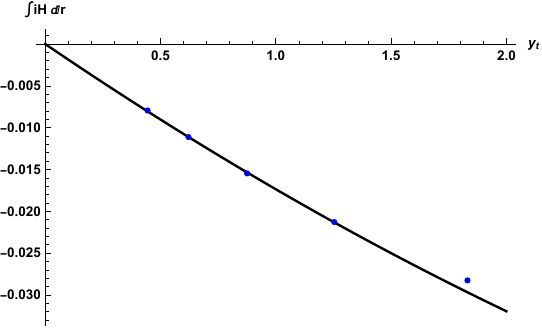} 
  \caption{Plot of $\int_0^{2h} dr\, iH$ vs.\ $\gf_t$ for the same times as in figure \ref{figE1}. The black line represents a fit by $-0.0187\, \gf_t + 0.00139\, \gf_t^2$ to the first four values.
  }
\label{figpintH}
\end{figure}

The function $2 i H=\Gtar$ has been plotted in figure \ref{figpGtr}.
It vanishes when $t\to 0$ for $r\neq h$ (as also mentioned in appendix \ref{appET}). To investigate the possibility of a finite remaining distribution at $r=h$, consider $I_H=\int_0^{2h} dr\, i H$. It turns out to be a non-linear function of $\ln(t)$ but an almost linear one as a function of $\gf_t=\gf(h,t)$; its four smallest values can be fitted by the form $\bt\, \gf_t + \gm\, \gf_t^2$, as shown in figure \ref{figpintH}.
Adding a constant $\al$ to the fit function leads to a rather small value $\al=0.00075$.
We assume that $H$ vanishes also as a distribution when $t\to 0$.\\

\bibliography{lit}

\end{document}